\documentclass[twocolumn,showpacs,preprintnumbers,nofootinbib]{revtex4-2}
\usepackage{color}
\usepackage{graphicx}
\usepackage{bm}
\usepackage{here}
\usepackage{braket}

\newcommand\nn{\nonumber \\}

\newcommand\bb{{\bm b}}
\newcommand\bk{{\bm k}}
\newcommand\bp{{\bm p}}
\newcommand\bx{{\bm x}}
\newcommand\bX{{\bm X}}

\newcommand\bn{{\bm n}}

\newcommand{\lk}{\left(}
\newcommand{\rk}{\right)}
\newcommand{\ltk}{\left\{}
\newcommand{\rtk}{\right\}}
\newcommand{\ldk}{\left[}

\newcommand{\rdk}{\right]}
\newcommand\beq{ \begin{eqnarray} }
\newcommand\eeq{ \end{eqnarray} }

\begin{document}
\title{Dissipation-relaxation dynamics of a spin-$1/2$ particle 
with a Rashba-type spin-orbit coupling in an ohmic heat bath}
\author{Tomohiro Hata}
\email{b19d6a08s@kochi-u.ac.jp}
\affiliation{Department of Mathematics and Physics, Kochi University, Kochi, Japan}
\author{Eiji Nakano}
\email{e.nakano@kochi-u.ac.jp}
\affiliation{Department of Mathematics and Physics, Kochi University, Kochi, Japan}
\author{Hiroyuki Tajima}
\email{hiroyuki.tajima@riken.jp}
\affiliation{Department of Physics, University of Tokyo, Tokyo, Japan}
\author{Kei Iida}
\email{iida@kochi-u.ac.jp}
\affiliation{Department of Mathematics and Physics, Kochi University, Kochi, Japan}
\author{Junichi Takahashi}
\email{takahashi.j@aoni.waseda.jp}
\affiliation{Faculty of Science and Engineering, Waseda University, Tokyo, Japan}
\begin{abstract}
Spin-orbit coupling (SOC), which is inherent to a Dirac particle that moves
under the influence of electromagnetic fields, manifests itself in a variety of
physical systems including non-relativistic ones.
For instance, it plays an essential role in spintronics developed in the past few decades,
particularly by controlling spin current generation and relaxation.
In the present work, by using an extended Caldeira-Leggett model,
we elucidate how the interplay between spin relaxation and momentum dissipation
of an open system of a single spin-$1/2$ particle with a Rashba type SOC is induced by the
interactions with a spinless, three-dimensional environment.
Staring from the path integral formulation for the reduced density matrix of the system,
we have derived a set of coupled nonlinear equations that consists of a quasi-classical Langevin
equation for the momentum with a frictional term and a spin precession equation.
The spin precesses around the effective magnetic field generated by
both the SOC and the frictional term.
It is found from analytical and numerical solutions to these equations that
a spin torque effect included in the effective magnetic field causes a spin relaxation
and that the spin and momentum orientations after a long time evolution are
largely controlled by the Rashba coupling strength. Such a spin relaxation mechanism
is qualitatively different from, e.g., the one encountered in semiconductors
where essentially no momentum dissipation occurs due to the Pauli blocking.

\end{abstract}


\maketitle

\section{Introduction}
Spin-orbit coupling (SOC) is ubiquitous in physics, ranging from
atomic fine structure \cite{landau_lifshitz_nonrelaquantum} to nuclear shell structure
\cite{On_Closed_Shells,On_the_Magic_Numbers},
spin dynamics in semiconductors \cite{electro_optic_modulator}, etc. 
The SOC originates from the $\mathcal{O}\lk m^{-2}\rk$ correction
in non-relativistic reduction of a charged Dirac particle of mass $m$
under electromagnetic fields \cite{Foldy_Wouthuysen};
e.g., for an electron of mass $m_e$, it reads
\beq
  H_{\rm so} = \frac{1}{2} \hat{\bm{\sigma}} \cdot \lk  \bm{\alpha} \times \hat{\bm{p}} \rk,
\eeq
where $\hat{\bm{\sigma}}$ are the Pauli matrices,
$\bm{\alpha} =-\frac{e\hbar}{2m_e^2 c^2}\bm{E}$ with $\bm{E}=-\bm{\nabla} \phi$ is an external electric field from a static potential $\phi$,
and no magnetic field is applied.
For simplicity, we will hereafter use natural units where $\hbar=1$.
%
For the past few decades, study of electronics pertaining to electron spin currents, i.e., spintronics,
has developed significantly for possible application to novel devices of information technology \cite{Interfacial_charge_spin_coupling},
where the spin Hall effect due to the SOC plays a crucial role in controlling the spin currents \cite{sinova}.
The platform for such devices is provided by semiconductors, for instance,
GaAs in which the transition between electrons in conducting and valence bands
leads to an effective mass $m^*$ that is much smaller than $m_e$ and hence
enhances the SOC at a level that cannot be ignored in comparison to that in vacuum \cite{rashba_semiconductor}.
In such semiconductors, the conducting electrons are moving effectively in a quasi-2 dimensional well confined in $z$ direction,
and two types of the SOC can be realized:
one is the Rashba type $H_{\rm so} \simeq \sigma_x p_y - \sigma_y p_x$ \cite{rashba_socoupling},
and the other the Dresselhaus type $H_{\rm so} \simeq \sigma_x p_x - \sigma_y p_y$ \cite{dresselhaus}.
Obviously, the SOC constitutes a part of the single-body Hamiltonian and leads to an energy splitting in spin states
given a finite momentum,
in addition to the Zeeman splitting by an external magnetic field
\cite{so-coupling_zeemaneffect},
and/or the Landau splitting by a spatial rotation \cite{so-coupling_rotationaleffect}.
Such a spin state, however, does not last long  in a coherent manner,
but relaxes to a lower energy state in the presence of various interactions with environmental degrees of freedom.
Thus, to know the spin relaxation process is important particularly for spintronics.


Recently, the SOC in cold atomic many-body systems has also attracted much attention.
A well-designed laser geometry provides these systems with artificial electromagnetic fields,
which act on the hyperfine states of trapped atoms and help
a pair of such pseudo-spin states with different momenta to couple with each other
so as to mimic the SOC of spin-$1/2$ electrons \cite{Stanescu_Galitski_2007}.
This technique was applied to experimental realization
of the Bose-Einstein condensation of spin-orbit coupled {\it bosonic} atoms
\cite{SOBEC}.
More interestingly, such atoms with or without the SOC can be confined
in anisotropic traps as minority (impurity) atoms, together with majority atoms as an environment,
which simulates a polaron problem for atomic impurities.
This kind of atomic impurities, referred to as Fermi or Bose  polarons according to
whether the majority atoms are fermions or bosons,
has been observed in cold atomic experiments \cite{PhysRevLett.102.230402,PhysRevLett.117.055301,PhysRevLett.117.055302}.
\par
Theoretically, such an atomic impurity without the SOC has been treated as a quantum open system,
where the dissipation of energy and momentum of the impurity occurs due to the interaction
with a cold or hot bosonic environment
\cite{Lampo,Boyanovsky_2019,2019NJPh...21d3014K}.
In the presence of the SOC, however, the interplay between momentum dissipation and spin relaxation
of the impurity has yet to be investigated.
Here it should be noted that the dissipative dynamics of a mobile atomic impurity
is different from the electron-spin relaxation dynamics in semiconductors in the sense
that the latter involves essentially no momentum dissipation.
Indeed, the Pauli blocking only allows
the electron momentum to change, during scattering processes with environmental degrees of freedom like phonons,
from $\bk$ to $\bk'$ just on the Fermi surface, i.e., $|\bk|=|\bk'|=k_F$ with the Fermi momentum $k_F$,
which leads to no dissipation.
More intriguingly,
such momentum changing processes cause effective magnetic-field fluctuations
to act on the spin degrees of freedom through the SOC in such a way as to
relax the spin orientation towards a possible lower-lying energy state \cite{Dyakonov1971},
as in nuclear magnetic resonances
where effective magnetic-field fluctuations are provided by environmental electron spins
\cite{Bloembergen_NMR,kubo_tomita_NMR,Solomon_spinrelaxation}.

In the present study
we demonstrate the spin relaxation of a spin-$1/2$ particle,
which occurs together with the momentum dissipation,
by employing an extended version of the one-dimensional Caldeira-Leggett (CL) model \cite{Caldeira1983}
in such a way as to be applicable to the particle that moves with a Rashba type SOC in a three dimensional environment.
In particular, we figure out a possible mechanism for the interplay between the spin relaxation and momentum dissipation
by simultaneously analyzing quasi-classical equations for the spin orientation and for the momentum.
%
The CL model is known to derive the Langevin equation for a quantum Brownian particle:
Starting from the von Neumann equation for the full density matrix and integrating
out the environmental degrees of freedom,
one can read off the Langevin equation from the resulting effective action
in the path integral formulation of the reduced density matrix of the particle.
This equation inevitably demonstrates a breakdown of the unitary evolution of
such an open system.
There is, however, a caveat in the use of the CL model: The positivity of the reduced density matrix
in the CL master equation is violated
in a short timescale even at high temperature \cite{Ambegaokar,DIOSI1993517,Schlosshauer}.
In this study, therefore, we assume that the environment's temperature and the timescale after decoherence are
sufficiently high and long, respectively, for us to restrict ourselves to a quasi-classical regime of the
Langevin dynamics \cite{schmid},
instead of evaluating directly the time evolution of the reduced density matrix
by employing, e.g., empirical Lindblad forms  \cite{lindblad,DIOSI1993517,Gao_PhysRevLett.79.3101}
that circumvent the positivity violation.

The remaining sections are organized as follows:
In Sec.~II we give a model Hamiltonian that consists of three parts, namely,
a single spin-$1/2$ particle system with a Rashba type SOC,
an environment of many-body harmonic oscillators,
and the interaction between the particle and the environment.
We then evaluate the effective propagator for the reduced density matrix of the particle
within the path-integral influence-functional method by Feynman and Vernon \cite{Feynman},
in which we introduce the spin coherent state for the path-integral representation of the spin degrees of freedom.
We finally derive quasi-classical dynamical equations for the particle's spin degrees of freedom and momentum
from the effective action in the path-integral formulation.
In Sec.~III we present numerical simulations for the spin relaxation and momentum dissipation,
which in turn are classified into typical patterns of the dynamics according to the model parameters that
govern the relaxation and dissipation (damping) time scales.
Section IV is devoted to summary and outlooks.
\section{Formulation}

In this section, we present the system-plus-environment-plus-interaction Hamiltonian, the eigen energy of the system,
the Feynman-Vernon influence functional, the quasi-classical dynamical equations, and the asymptotic state of the system.

\subsection{A model Hamiltonian}
We consider a system-plus-environment-plus-interaction model described
by the Hamiltonian $\hat{H} = \hat{H}_S + \hat{H}_B + \hat{H}_I$,
where
\beq
  \hat{H}_S &=& \frac{\hat{\bm{p}}^2}{2m}
  + \hat{\bm{s}} \cdot (\bm{\alpha} \times \hat{\bm{p}} + \bm{B}), \\
  \hat{H}_B &=& \frac{1}{2}  \sum^{\infty}_{k=0} \left[ \hat{\bm{P}}^2_{k} + \omega^2_k \hat{\bm{X}}^2_{k} \right], \\
  \hat{H}_I &=& - \hat{\bm{x}} \cdot \lk \sum^{\infty}_{k=0} c_k  \hat{\bm{X}}_{k}\rk.
\eeq
$\hat{H}_S$ denotes the Hamiltonian of the system of a spin-$1/2$ particle moving with a Rashba type SOC,
which is characterized by the spin operator
$\hat{\bm{s}}=\frac{1}{2}\hat{\bm{\sigma}}$,
the momentum operator $\hat{\bm{p}}$,
and a constant vector $\bm{\alpha}=(0,0,\alpha)$ whose size determines the Rashba coupling strength.
In addition, we assume that an external field $\bm{B}$ brings about
the Zeeman term $\hat{\bm{s}}\cdot \bm{B}$,
which appears in general for particles having a nonzero spin and an intrinsic dipole magnetic moment parallel to the spin.
Note that in the present study we employ units in which the size of the dipole magnetic moment is unity.
We also assume throughout the present study
that $\bm{B}$ is parallel to the $z$ axis, i.e., $\bm{\alpha}$, as in semiconductor experiments,
and that the $z$ component of the particle's momentum is always zero.
$\hat{H}_B$ is the Hamiltonian of the environment composed of an infinite number of harmonic oscillators,
which is characterized by the angular frequency $\omega_k$, momentum operator $\hat{\bm{P}}_{k}$,
and coordinate operator $\hat{\bm{X}}_{k}$ of each mode $k$.
$\hat{H}_I$ describes the interaction between the system and the environment, which is characterized
by the linear coupling between their coordinate operators, i.e., $\hat{\bm{x}} \cdot \hat{\bm{X}}_k$,
with the strength $c_k$ for each mode $k$.
The operators satisfy the canonical relations
$\ldk \hat{x}_i, \hat{p}_j\rdk=i\delta_{ij}$ and $\ldk \hat{X}_{k,i}, \hat{P}_{k',j}\rdk=i\delta_{ij}\delta_{k k'}$.
It should be noted that $\hat{H}_B$ together
with $\hat{H}_I$ constitutes the Caldeira-Leggett type heat bath model.

\subsection{Single particle energies}
\begin{figure}[h]
    \includegraphics[width=1.0\linewidth]{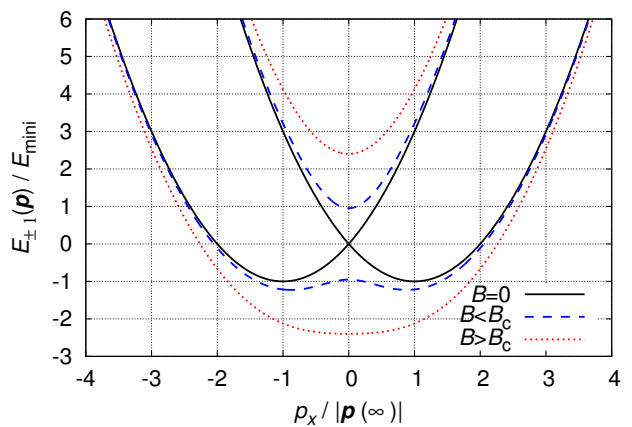}
      \caption{Single particle energies $E_{\pm1}(\bp)$ as functions of $p_x$
      for $B=0$ (solid lines), $B<B_{\rm c}$ (dashed), and $B>B_{\rm c}$ (dotted).
      The upper (lower) lines correspond to $E_{+1}$ ($E_{-1}$).
      Note that these energies have the rotation symmetry in $p_x$-$p_y$ plane,
      while the plotted values of the momenta and energies are normalized, respectively,
      by the asymptotic value (\ref{aysm11}) at $B=0$ and the absolute value of
      $E_{\rm min}=-m\alpha^2/8$ corresponding to $E_{-1}(\bm{p}(\infty))$ at $B=0$.}
    \label{fig1}
\end{figure}

In order to clarify conserved quantities of the system,
which eventually undergo dissipation and relaxation under the influence of the environment,
we first obtain the solution of the eigen value problem solely for the system's Hamiltonian as
\beq
\hat{H}_S |\bp,s\rangle &=& E_s(\bp) |\bp,s\rangle,
\label{eigen1}
\eeq
with
\beq
E_s(\bp) &=& \frac{\bp^2}{2m}+s \frac{1}{2}|\bm{\alpha} \times \bp + \bm{B}|,
\label{eigen2}
\eeq
where $|\bp,s\rangle$ represents the eigen state with the eigen values of the momentum $\bp$
and the spin doublet $s=\pm 1$ with respect to the quantization axis parallel
to $\bm{\alpha} \times \bm{p} + \bm{B}$.
The system particle thus keeps having a constant momentum $\bp$ once given,
and its spin expectation value precesses about the constant vector $\bm{\alpha} \times \bm{p} + \bm{B}$,
if there is no influence from the environment.
Note that since we have taken $\bm{B}=(0,0,B)$,
$\hat{H}_S$ possesses the rotational symmetry about the $z$ axis.
The single particle energies are depicted in Fig.~\ref{fig1} for some characteristic values of $B$;
there exists a critical value of the magnetic field,
\beq
B_{\rm c}=m\alpha^2/2,
\label{critB1}
\eeq
above which the degenerate minima of $E_{-1}$ merge into one.

\subsection{Feynman-Vernon influence functional}
We now proceed to consider the time evolution of the system's density matrix
$\rho_t^S$ (the reduced density matrix) under the influence of the environment.
To this end, we employ the path integral formalism,
from which the effective action of the system
and the corresponding quasi-classical dynamical equations can be exploited.

We start with the time evolution of the full density matrix $\rho_t$,
which is governed by the von Neumann equation,
\beq
  \frac{d \hat{\rho}_t}{dt} &=& -i [\hat{H}, \hat{\rho}_t],
\eeq
with
\beq
  \hat{\rho}_t &=& \exp{(-i\hat{H}t)} \hat{\rho}_0 \exp{(i\hat{H}t)},
\label{dm2}
\eeq
where $\hat{\rho}_0$ represents the initial density matrix at $t=0$.
We assume that $\hat{\rho}_0$
is given by the direct product of the initial density matrices of the system and environment as
\beq
  \hat{\rho}_0= \hat{\rho}^S_0 \otimes \hat{\rho}^B_0,
\eeq
where the environmental part is in thermal equilibrium of temperature $T$, i.e.,
$\hat{\rho}^B_0=e^{-\beta \hat{H}_B}/Z$ with $\beta=1/k_B T$.

The element of the reduced density matrix with respect to
the particle's coordinate $\bx$ and spin state $g$ is given
by taking a trace over the environmental coordinates as
\beq
\rho_t^S(\bx g, \bx' g') &=& \int_{\bX} \langle \bx,g; \bX | \hat{\rho}_t |\bx',g'; \bX \rangle,
\label{rd1}
\eeq
where $\int_{\bX}\equiv \Pi_k  \int {\rm d}{\bX}_k$,
and we represent the spin state by a general $SU(2)$ rotation of the highest weight spin eigen state  \cite{altland_simons} as
\beq
  \ket{g}
  &=& e^{-i\phi \hat{s}_3} e^{-i\theta \hat{s}_2} e^{-i\psi \hat{s}_3} \ket{+1}
\label{scs}
\eeq
with the Euler angles $\phi$, $\theta$, and $\psi$.
This representation is useful to express the reduced density matrix (\ref{rd1})
in the path integral formalism,
where the path integral coordinates of the spin state can be represented
by continuous compact parameters, i.e., the Euler angles, on the $S^3\sim SU(2)$ manifold,
and the decomposition of the identity
used in the path integral is given in terms of the Haar measure in $SU(2)$ as
\beq
  \int_{S^3} {\rm d}g \, |g\rangle \langle g| ={\rm \bm{I}},
\eeq
where $\int_{S^3}{\rm d}g=\frac{1}{8\pi^2} \int^{\pi}_0 \sin\theta {\rm d}\theta \int^{2\pi}_0 {\rm d}\phi \int^{4\pi}_0{\rm d}\psi $, and ${\rm \bm{I}}$ is the identity operator.
For the spin coherent state (\ref{scs}), the spin expectation value can be expressed in terms of the Bloch sphere coordinates as
\beq
  \bn &\equiv& \bra{g} \hat{\bm{s}} \ket{g}
  = \frac{1}{2} (\sin \theta \cos \phi , \sin \theta \sin \phi , \cos \theta ).
  \label{Bloch1}
\eeq
The remaining angle $\psi$
is hidden in $|g\rangle$ as an overall $U(1)$ gauge factor.
Thus we will denote the Bloch sphere coordinates simply by
\beq
g =\ltk \theta, \phi \rtk,\ g' = \ltk \theta', \phi' \rtk,
\eeq
etc.

In the Feynman-Vernon influence functional method \cite{Feynman},
the time evolution of the reduced density matrix element can be expressed
in terms of the propagator $G_t$ from the initial state as
\begin{equation}
  \rho^S _t(\bm{x} g, \bm{x}' g')
  = \int
  _{\bar{\bx},\bar{g},\bar{\bx}',\bar{g}'}
  G_t(\bx g, \bx' g'; \bar{\bx} \bar{g}, \bx' \bar{g}') \, \rho^S _0(\bar{\bx}\bar{g}, \bx'\bar{g}'),
\end{equation}
where
$\rho^S _0(\bar{\bx}\bar{g}, \bx'\bar{g}')
= \langle \bar{\bx},\bar{g} | \hat{\rho}_0^S |\bx',\bar{g}'\rangle$,
and the propagator has the forward-backward  path integral representation
\beq
&&G_t(\bx g, \bx' g'; \bar{\bx} \bar{g}, \bar{\bx}' \bar{g}')
\nn
&=& \int_{\bX, \bar{\bX}, \bar{\bX}'} 
\bra{\bx,g; \bX} e^{-i\hat{H}t} \ket{\bar{\bx},\bar{g}; \bar{\bX}} \, \rho^B\lk \bar{\bX}, \bar{\bX}'\rk
\nn && \qquad  \qquad \qquad \times
\bra{\bar{\bx}',\bar{g}'; \bar{\bX}'} e^{i\hat{H}t} \ket{\bx',g';\bX}
\\
&=&
\int \mathcal{D}\bx \mathcal{D}\bx' \mathcal{D}g \mathcal{D}g' \, e^{i W\ldk \bx g, \bx'g'\rdk},
\eeq
where $\rho^B( \bar{\bm{X}},\bar{\bm{X}}^{\prime})= \langle \bar{\bm{X}} | \hat{\rho}_0^B |\bar{\bm{X}}^{\prime}\rangle$, and
the effective action of the system is given by
\beq
  W\ldk \bx g, \bx' g'\rdk &=&
  \mathcal{A}^S\ldk \bx g\rdk - \mathcal{A}^S\ldk \bx' g'\rdk
  \nn &&
  + W_1\ldk \bx, \bx' \rdk + W_2\ldk \bx, \bx' \rdk.
  \label{effact1}
\eeq
Here, $\mathcal{A}^S$ is the action of the particle alone,
\beq
\mathcal{A}^S\ldk\bx g\rdk&=& \int^t_0{\rm d}u\, L^S\ldk \bx(u), g(u)\rdk
\label{Sact1}
\eeq
with
\beq
L^S &=& 
\frac{m}{2}\lk\dot{\bm{ x}} - \bm{n}\times \bm{\alpha}\rk^2
+\frac{1}{2}  \dot{\phi} \cos \theta -\bn \cdot \bm{B},
\label{Sact2}
\eeq
which satisfies
the boundary conditions at the initial time, i.e., $u=0$,
\beq
\ltk \bx(0),g(0) \rtk = \ltk \bar{\bx}, \bar{g} \rtk,
\ \ltk \bx'(0), g'(0)\rtk = \ltk  \bar{\bx}', \bar{g}'\rtk,
\eeq
and at the final time, i.e., $u=t$,
\beq
\ltk \bx(t), g(t)\rtk = \ltk \bx, g \rtk, \ \ltk \bx'(t), g'(t)\rtk = \ltk \bx', g'\rtk.
\eeq
Note that the dot in Eq.\ (\ref{Sact2}) denotes the time derivative.
Also,
$W_1$ and $W_2$ are respectively the imaginary and real parts of the influence functional,
i.e., the contribution to the effective action from $H_B$ and $H_I$, as given by
\beq
W_1\ldk \bx, \bx' \rdk
&=& i \int^t_0 {\rm d}u \int^u_0 {\rm d}u^{\prime } \int^{\infty}_0 {\rm d}\omega \,
J(\omega)\, \coth{\frac{\omega \beta}{2} }
\nn
&& \quad \times \, \cos{\omega (u-u')}\,
\bm{x}_-(u) \cdot \bm{x}_-(u'),
\nn
\label{w1}
\\
W_2\ldk \bx, \bx' \rdk
&=&
2 \int^t_0 {\rm d}u \int^u_0 {\rm d}u^{\prime} \int^{\infty}_0 {\rm d}\omega\, J(\omega)
\nn
&& \quad \times \sin\omega (u-u') \, \bm{x}_-(u) \cdot \bm{x}_+(u^{\prime}),
\label{w2}
\eeq
where
\beq
  J(\omega) &=& \sum_k \frac{c^2_k}{2 \omega_k} \delta (\omega - \omega_k)
\label{SDF}
\eeq
is the spectral density function, and
\beq
\bx_+(u) &=& \frac{\bx(u) +\bx'(u)}{2}, \\
\bx_-(u) &=& \bm{x}(u) - \bx'(u)
\eeq
represent respectively the center of mass and relative coordinates
with respect to the forward and backward path integrals,
i.e., the diagonal and off-diagonal elements of the reduced density matrix
$\rho_t^S(\bx g, \bx' g')$ at each $u$ with $0\leq u\leq t$.

\subsection{Quasi-classical dynamical equation}
From here on, as in the Caldeira-Leggett model,
we assume that spectral density is ohmic, i.e.,
\beq
  J(\omega) = \frac{c\, \omega}{\pi},
\eeq
where $c$ is a constant,
\footnote{
It should be noted that in the non-ohmic case, i.e.,  $J(\omega)\sim \omega^\alpha$ with $\alpha \neq 1$,
the non-locality in $W_2$ is maintained so that the memory effect comes out in the resultant Langevin equation.
}
and then we obtain
\beq
W_2\ldk \bm{x}, \bm{x}^{\prime}\rdk
&=& -
c\int^t_0 {\rm d}u \, \dot{\bm{x}}_+(u) \cdot \bm{x}_-(u)
  \nn
&&
+c'
\int^t_0 {\rm d}u \, \bm{x}_+(u) \cdot \bm{x}_-(u)
  \nn
&&
-
c   
\bx_+(0)\cdot\bx_-(0),
\label{W2eq}
\eeq
where the coefficient in the second term is interpreted as $c'=\sum_k \frac{c^2_k}{\omega_k^2}$.
The last two terms will be ignored hereafter
because the second term can be renormalized into an external potential of the particle,
which we suppose to be absent in the present study
\footnote{The second term, which comes from the original form
(\ref{SDF}) of the spectral function, would add an elastic force with
negative spring constant and hence acts to destabilize the system,
but one could counteract such a possible instability by adding an
harmonic trap potential.}, and the last one is a surface term irrelevant for the dynamics.


Here we also assume that the environment temperature is sufficiently
higher than the excitation energies,
i.e., $k_B T \gg  \omega_k$, so that $W_1$ can be approximated by
\cite{schmid}
\beq
W_1\ldk \bm{x} , \bm{x}^{\prime}\rdk &\simeq&
i
c k_B T \int^t_0 {\rm d}u \, \bx^2_-(u).
\label{w1hta1}
\eeq
This approximation is consistent with the quasi-classical description that we will develop below.


Under these assumptions just made above,
let us now take the optimal condition for the effective action, i.e., $\delta W=0$,
with respect to the off-diagonal variables $\bx_-$, $\theta_-=\theta -\theta'$, and $\phi_-=\phi-\phi'$
at $\bx_-=\theta_-=\phi_-=0$
to obtain the equations for the quasi-classical (diagonal) variables:
\beq
  &&\ddot{\bx}_++ \gamma\, \dot{\bx}_+- (\dot{\bn}_+ \times \bm{\alpha}) = 0,
  \label{xeq1} \\
  &&\dot{\bn}_+ =
  \ldk \bm{\alpha} \times m\lk \dot{\bx}_+ - \bn_+ \times \bm{\alpha}\rk +\bm{B} \rdk \times \bn_+,
  \label{neq1}
\eeq
where $\gamma= c/m$ is the friction coefficient, 
and $\bn_+=\bn|_{\theta=\theta_+, \phi=\phi_+}$
with the diagonal Euler angles $\theta_+=\lk \theta+\theta'\rk/2$ and $\phi_+=\lk \phi+\phi'\rk/2$.
The first equation (\ref{xeq1}) corresponds to the quasi-classical dynamical equation
with the friction term $\gamma\, \dot{\bx}_+$
and with the additional term involving $\bm{\alpha}$ that comes from the SOC,
while the second equation (\ref{neq1}) mainly governs the spin dynamics
that can be seen as an instantaneous spin precession
about a velocity-dependent effective magnetic field.
The precession equation is equivalent to the one derived from the Heisenberg equation of motion
in terms of the canonical momentum as will be shown in Eq.\  (\ref{neq2}) below.


Here it should be noted that the optimal condition to derive Eqs.\ (\ref{xeq1}) and (\ref{neq1})
corresponds to tracing the diagonal path $\bx_+$ that is not affected by
off-diagonal fluctuations $\bx_-$ and $g_- \equiv g-g'$ at any $u$ that satisfies $0<u<t$.
In this sense these equations determine the quasi-classical trajectory of the particle, i.e.,
the wave packet of width $|\bx_-|$, which is in turn influenced systematically by its own quasi-classical
spin dynamics through the SOC, and vice versa.
This sort of classical description is justified at high temperatures,
because the time evolution of the reduced density matrix is dominated by the imaginary part (\ref{w1hta1}) as
\beq
    \partial_t \rho_t^S (\bm{x}g,\bm{x}^{\prime} g^{\prime})
    &\simeq&
    -2  
    \pi\gamma\, \left( \frac{\bm{x}_-}{\lambda_T} \right)^2
    \, \rho_t^S\lk \bx g,\bx' g'\rk,
\eeq
which implies that the probability of having a finite off-diagonal fluctuation of size $|\bx_-|=|\bx-\bx'|$
diminishes exponentially fast and the corresponding decoherence time scale
\beq
\tau_{\rm dec}= \lk
       2 
      \pi\gamma\rk^{-1} \, \left( \frac{\bm{x}_-}{\lambda_T} \right)^{-2}
\label{dect1}
\eeq
becomes smaller than the damping time scale $\gamma^{-1}=m/ c$
for sufficiently large width $|\bx_-|$ of the particle wave packet
compared with the thermal de~Broglie wavelength
$\lambda_T=\hbar\sqrt{2\pi/m k_{\rm B} T}$ \cite{Zurek,RevModPhys_zurek}.
Thus the decoherence of the superposition between different coordinates
is quickly achieved.
We can also expect a fast decoherence with respect to the spin variable $g_-$
as in the case of $\bx_-$.
Discussion of this point is given in appendix~A.


In order to make Eqs.\  (\ref{xeq1}) and (\ref{neq1}) more transparent,
it is instructive to introduce the canonical momentum of the particle,
which can be obtained from the Lagrangian (\ref{Sact2}) as
\beq
  \bm{p}
  &=& \frac{\partial L_S}{\partial \dot{\bm{x}}}
  = m \lk \dot{\bm{x}} - \bm{n}\times \bm{\alpha} \rk,
\label{canonical_p}
\eeq
and then to represent the equations as
\beq
  &&\dot{\bm{p}} + \gamma \bm{p}
  = m \gamma (\bm{\alpha} \times \bm{n}),
  \label{dotp_noxi} \\
  &&\dot{\bm{n}} = (\bm{\alpha} \times \bm{p} +\bm{B}) \times \bm{n},
  \label{neq2}
\eeq
where we have simplified the notation as $\bx_+\rightarrow \bx$ and $\bn_+\rightarrow \bn$.
The dynamics of $\bp$ is mainly controlled by Eq.~(\ref{dotp_noxi}),
which includes the inhomogeneous (source) term
attributable to both of the dissipation ($\gamma$) and SOC ($\bm {\alpha}$) effects.
The source term alters the orientation of $\bp$ during the momentum dissipation as
will be directly observed from numerical results in the next section.
On the other hand,
Eq.~(\ref{neq2}) leads to the spin precession
about the axis of $\bm{B}$ plus the effective magnetic field
defined by $\bm{b}_{\rm eff}= \bm{\alpha} \times \bm{p}$.
Since
$\bm{b}_{\rm eff}$ changes its direction and magnitude with time
in accordance with the dynamics of $\bp$,
the spin precession around $\bm{b}_{\rm eff}+\bm{B}$ is only instantaneous.
We can see this situation more explicitly
by rewriting the equations in terms of the the effective magnetic field as
\beq
  &&\dot{\bm{b}}_{\rm eff}
  + \gamma \, \bm{b}_{\rm eff}
  = m \gamma \, \bm{\alpha} \times \lk\bm{\alpha} \times \bm{n}\rk,
  \label{beffeq1}
  \\
  &&\dot{\bn} = \lk \bm{b}_{\rm eff} +\bm{B}\rk \times \bm{n}
  \label{dotn_beff}.
\eeq
In what follows, we will simultaneously solve the set of equations,
i.e., Eqs.\ (\ref{beffeq1}) and (\ref{dotn_beff}), and discuss
the dissipation-relaxation dynamics of the spin and momentum up to the
possible final state in a manner that depends on the parameters
$\bm{\alpha}$ and $\gamma$ as well as on the initial condition.

\begin{figure}[h]
 \includegraphics[width=1.0\linewidth]{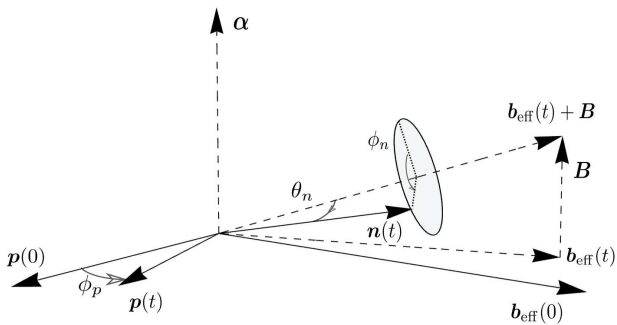}
     \caption{Schematic view of the three-dimensional vectors that represent the
     spin $\bn(t)$, the momentum $\bp(t)$, and the effective and external
     magnetic fields $\bb_{\rm eff}(t)$ and $\bm{B}$.
     The zenith and azimuth angles $\theta_n$ and $\phi_n$ of the spin direction with respect to the total magnetic field
     $\bb_{\rm eff}+\bm{B}$ are indicated. The angle between $\bp(t)$ and $\bp(0)$ is denoted by $\phi_p$.
     The Rashba coupling $\bm{\alpha}$ and initial momentum $\bp(0)$ are set in the $z$ and $x$ direction, respectively,
     while both of $\bp$ and $\bb_{\rm eff}$ always lie in $x$-$y$ plane. }
    \label{fig2}
\end{figure}

As a first step, for given $\bn(u)$,
we solve Eq.~(\ref{beffeq1}) with respect to $\bm{b}_{\rm eff}$ analytically
using the retarded Green function with the boundary condition at $u=0$ as
\beq
  \bm{b}_{\rm eff}(u)
  &=& \bm{b}_{\rm eff}(0)\, e^{-\gamma u}
  +m \gamma \int^{u}_0 {\rm d}u'\, e^{-\gamma (u-u') }
  \nn
  && \qquad \qquad  \
  \times \ldk \bm{\alpha} \cdot \bn(u') \, \bm{\alpha} -\bm{\alpha}^2\, \bn(u')\rdk.
  \label{beffsol1}
\eeq
Plugging the above result back to Eq.~(\ref{dotn_beff}),
we can observe that
the first term in the right side of Eq.~(\ref{beffsol1}) is responsible for the spin precession
around the direction of $\bm{b}_{\rm eff}(0)$ that lies on $x$-$y$ plane,
although it will damp exponentially.
The term proportional to $\bm{\alpha}$ in the square bracket in Eq.~(\ref{beffsol1}),
together with $\bm{B}$, also leads to the spin precession around $z$ axis.
In total, therefore, the axis of the spin precession at a given snapshot deviates
from 
$\bm{b}_{\rm eff}(0)$ axis, as illustrated in Fig.~\ref{fig2}.
More importantly,
the term proportional to $\bm{\alpha}^2 \bn(u)$ in Eq.\ (\ref{beffsol1}) plays the role of the spin torque
that will bend the spin towards the vertical direction
to both the past and present spin vectors, i.e., $\bn(t) \times \bn(u)$ for $t\ge u$ in Eq.\ (\ref{dotn_beff}).
The spin, therefore, eventually relaxes to the direction antiparallel to the spin quantization axis, i.e.,
$\bm{\alpha}\times \bp(\infty) +\bm{B}$, as the energetically favorable direction,
where the $\bp(\infty)$ is the asymptotic momentum that gives
one of the degenerate minima of the lower single particle energy $E_{-1}(\bp)$.

\subsection{Asymptotic behavior of the dynamical variables}
The fate of the dynamical system is characterized by the asymptotic behavior of the
variables, i.e., $\bn(\infty)$ and $\bp(\infty)$ (or, equivalently, $\bm{b}_{\rm eff}(\infty)$).
Such behavior
can be deduced from the static limit of Eqs.\ (\ref{dotp_noxi}) and (\ref{neq2}),
leading to
\beq
&& \bp = m\, \bm{\alpha} \times \bm{n},
\ \ \ldk \bm{\alpha} \times \bp + \bm{B} \rdk \times \bn =0
\nn
\rightarrow
&& \ldk m\lk \bm{\alpha} \cdot \bm{n}\, \bm{\alpha} -\bm{\alpha}^2\, \bn\rk
+ \bm{B} \rdk \parallel  \bn.
\label{asympB1}
\eeq


In the case of our interest,
where $\bm{B}=\lk 0, 0, B\rk$ and $0\le B < B_{\rm c}$,
the momentum becomes asymptotically vertical to the spin as
\beq
 \bp(\infty) &=& m\, \bm{\alpha} \times \bn(\infty),
 \label{asym1}
\eeq
and its magnitude can be obtained from the extreme condition $\partial_{\bp} E_{-1}=0$ as
\beq
|\bp(\infty)| = \frac{1}{\alpha} \sqrt{B_{\rm c}^2-B^2}.
\label{aysm11}
\eeq
The extreme condition above implies that the group velocity vanishes in the end, i.e., $\dot{\bx}(\infty)=0$;
see Fig.~\ref{fig11} in appendix~B.
Here it should be reminded again that $\bp$ is always confined in $x$-$y$ plane during the evolution.
The asymptotic spin, on the other hand,
has nonzero parallel and vertical components to $z$ axis when $B \neq 0$.
The $z$ component of the spin can be determined from the condition (\ref{asympB1})
as
\beq
&& m\, \bm{\alpha} \cdot \bm{n}(\infty)\, \bm{\alpha} + \bm{B} = 0
\nn
\rightarrow && n_z(\infty) = -\frac{1}{2} \frac{B}{B_{\rm c}},
\label{asym2}
\eeq
while the magnitude of the spin projection on $x$-$y$ plane
can be determined from $\bn^2=1/4$.
Note that the asymptotic behavior at zero magnetic field can be obtained
simply by taking the limit of $B\rightarrow 0$ in the above results.
In this case, the asymptotic spin has no $z$ component.
Incidentally, in the case of $B>B_{\rm c}$,
we obtain $\bp(\infty)=0$ as the minimum of the single particle energy $E_{-1}$ (see Fig.~\ref{fig1}),
and the spin eventually gets antiparallel to $\bm{B}$,  i.e., $n_z(\infty) =-1/2$.


The above analysis shows
that only the relative angles among $\bm{\alpha}$, $\bp$, and $\bn$ are fixed asymptotically
irrespective of the initial condition.
Since the Hamiltonian has the rotational symmetry around $z$ axis,
the direction of $\bp(\infty)$ on $x$-$y$ plane is determined by $\bp(0)$ and $\bn(0)$.
We will demonstrate this situation
by numerical simulations below and then classify the resultant asymptotic behavior
into two characteristic cases.

\section{Numerical simulations and discussion}
Before going into detailed calculations,
we define the initial precession period
and the damping time as
\beq
\tau_{\rm prec} &=& \frac{2\pi}{\sqrt{\alpha^2 \bp^2(0) +B^2}},
\label{prect1}
\\
\tau_{\rm damp} &=& \gamma^{-1} = m/
c,
\label{dampt1}
\eeq
both of which are assumed much longer than the decoherence time scale (\ref{dect1})
for a typical size of the particle wave packet.
In numerical calculations we will use them as the reference time scales
for classification of dynamical domains.


We always set the initial momentum parallel to $x$ axis as $\bm{p}(0) = (p_0, 0, 0)$ without loss of generality
and then observe the time evolution of the variables
to see how their final state depends on the initial condition of $\bn(0)$
for some typical values of $\alpha$ and $\gamma$ as well as a fixed value of $p_0$.
The spin direction is specified by the angles $\theta_n$ and $\phi_n$, as depicted in Fig.~\ref{fig2}:
The zenith axis is given along the instantaneous precession axis $\bb_{\rm eff} +\bm{B}$,
the zenith angle $\theta_n$ is taken between $\bb_{\rm eff} +\bm{B}$ and $\bn$,
and the azimuthal angle $\phi_n$ between $\bn$ and $\bm{\alpha}$ projected
on the plane normal to $\bb_{\rm eff} +\bm{B}$.
The angle between $\bp(t)$ and $\bp(0)$ is denoted by $\phi_p$.

\subsection{Zero magnetic field: $B=0$}
\begin{figure}[h]
    \begin{tabular}{r}
      \includegraphics[width=1.0\linewidth]{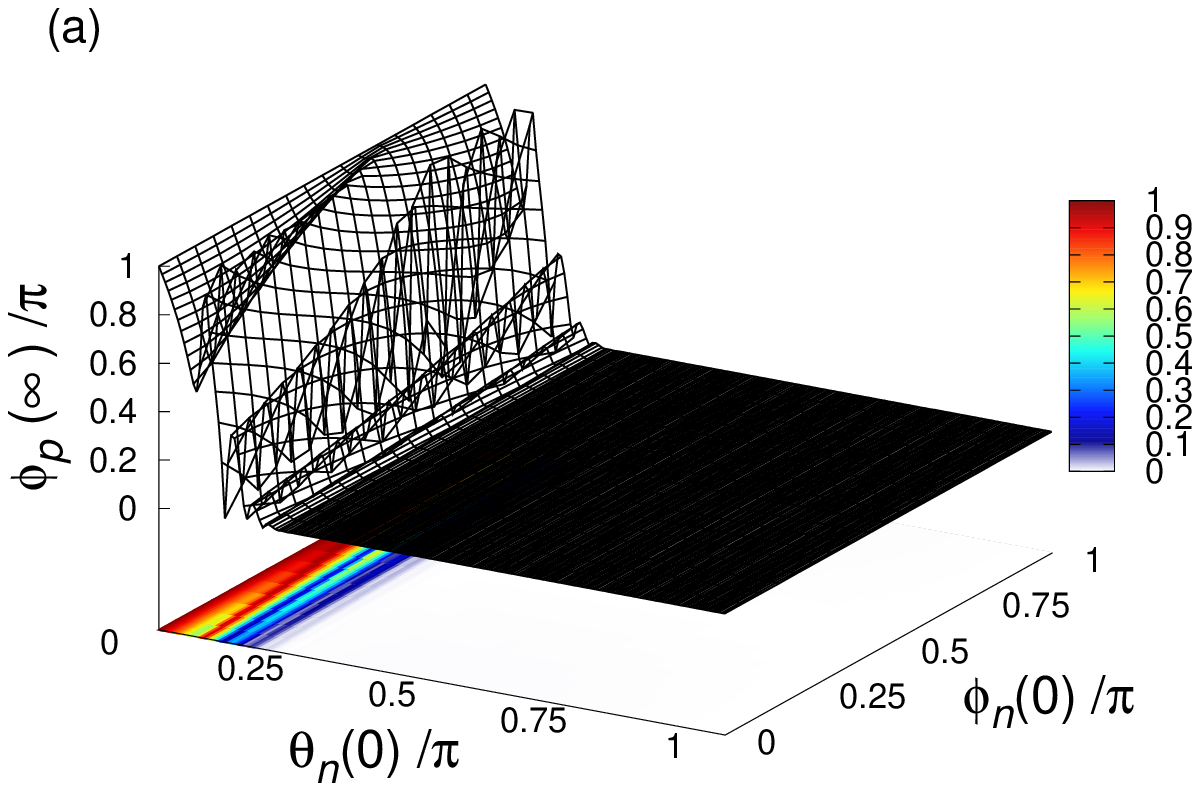} \\
      \includegraphics[width=1.0\linewidth]{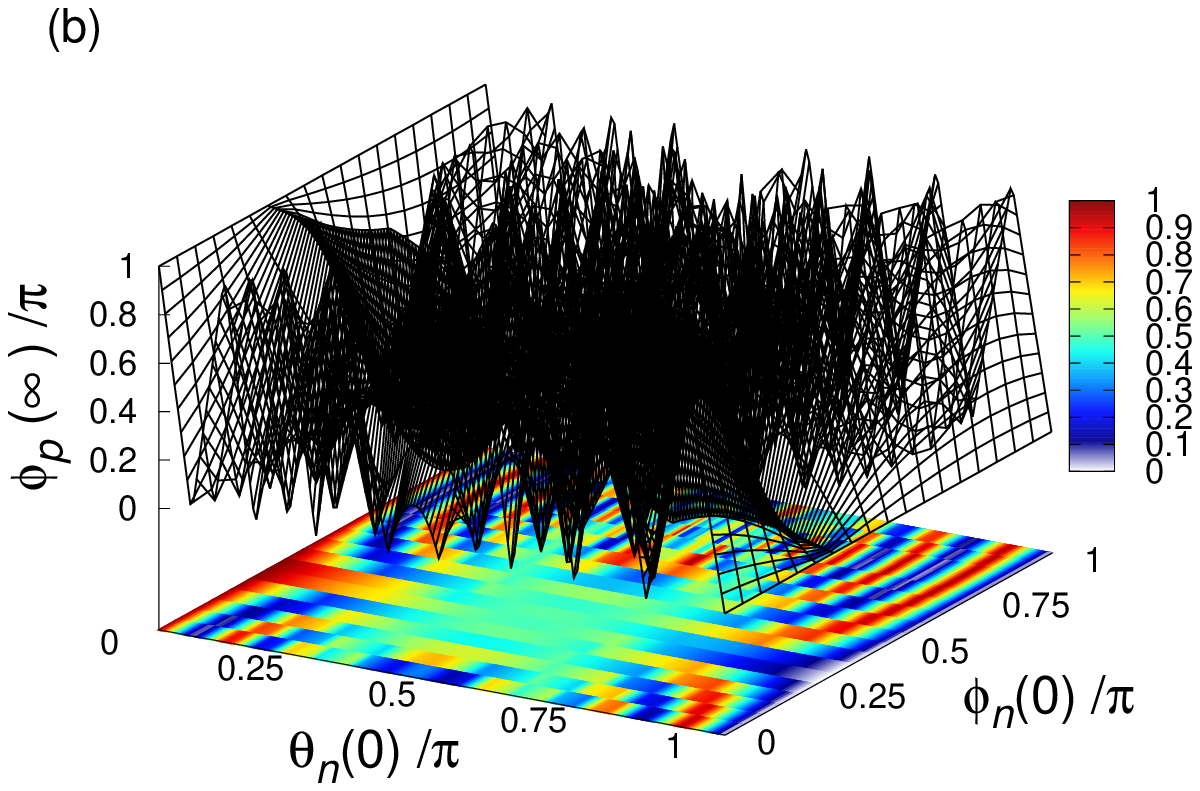}
    \end{tabular}
      \caption{Asymptotic values of the momentum angle $\phi_p(\infty)$ as functions of the initial values of the spin angles $\theta_n(0)$ and $\phi_n(0)$,
      plotted for $\tau_{\rm prec} \ll \tau_{\rm damp}$ (a) and $\tau_{\rm prec} \gg \tau_{\rm damp}$ (b) at $B=0$.
      The parameters are set as follows:
      $\alpha p_0/\gamma=167.50$ (a) and $\alpha p_0/\gamma=3.6750$ (b), as well as $m \gamma /p^2_0 =0.00081633$ in both cases.}
    \label{fig3}
\end{figure}

We first examine the case of $B=0$ numerically.
In Fig.~\ref{fig3} we show the result for the asymptotic value of the momentum angle $\phi_p(\infty)$
as functions of the initial values of the spin angles $\theta_n(0)$ and $\phi_n(0)$
both for $\tau_{\rm prec} \gg \tau_{\rm damp}$ and $\tau_{\rm prec} \ll \tau_{\rm damp}$.
Since we have fixed the initial values of $p_0$ and $\gamma$ to some specific values,
the precession period $\tau_{\rm prec}$ is determined solely by the strength of the Rashba coupling $\alpha$.
Figure~\ref{fig3}(a) shows that for strong Rashba coupling, that is, for $\tau_{\rm prec}\ll \tau_{\rm damp}$,
the asymptotic momentum direction does not deviate so much from its initial one
for a wide range of the initial spin angles satisfying $\theta_n(0) \gtrsim \pi/4$.
For weak Rashba coupling, that is, for $\tau_{\rm prec}\gg \tau_{\rm damp}$, on the other hand,
the asymptotic direction of the momentum fluctuates significantly as the initial spin angles change only slightly.
This feature can be seen clearly from Fig.~\ref{fig3}(b).

\begin{figure*}[htbp]
  \centering
    \begin{tabular}{cc}
      \includegraphics[width=0.5\linewidth]{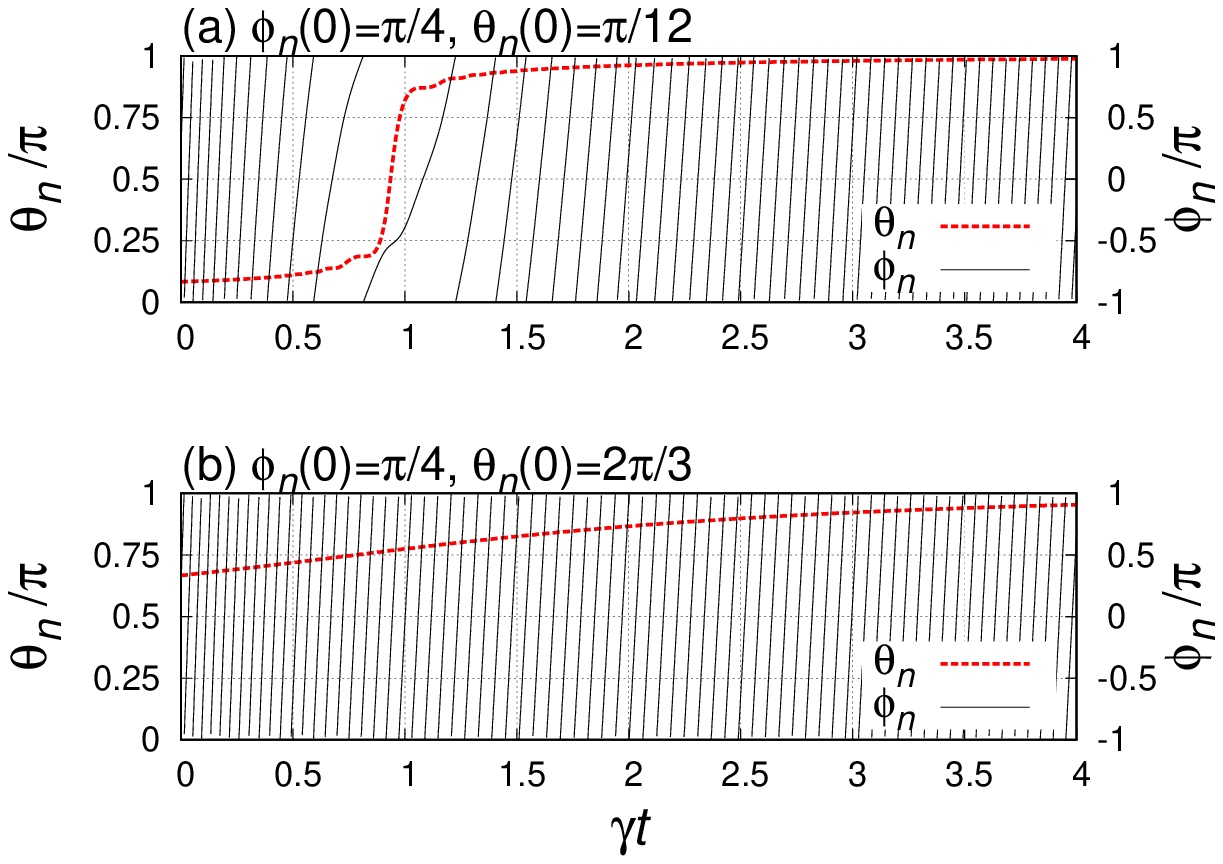} &
      \includegraphics[width=0.5\linewidth]{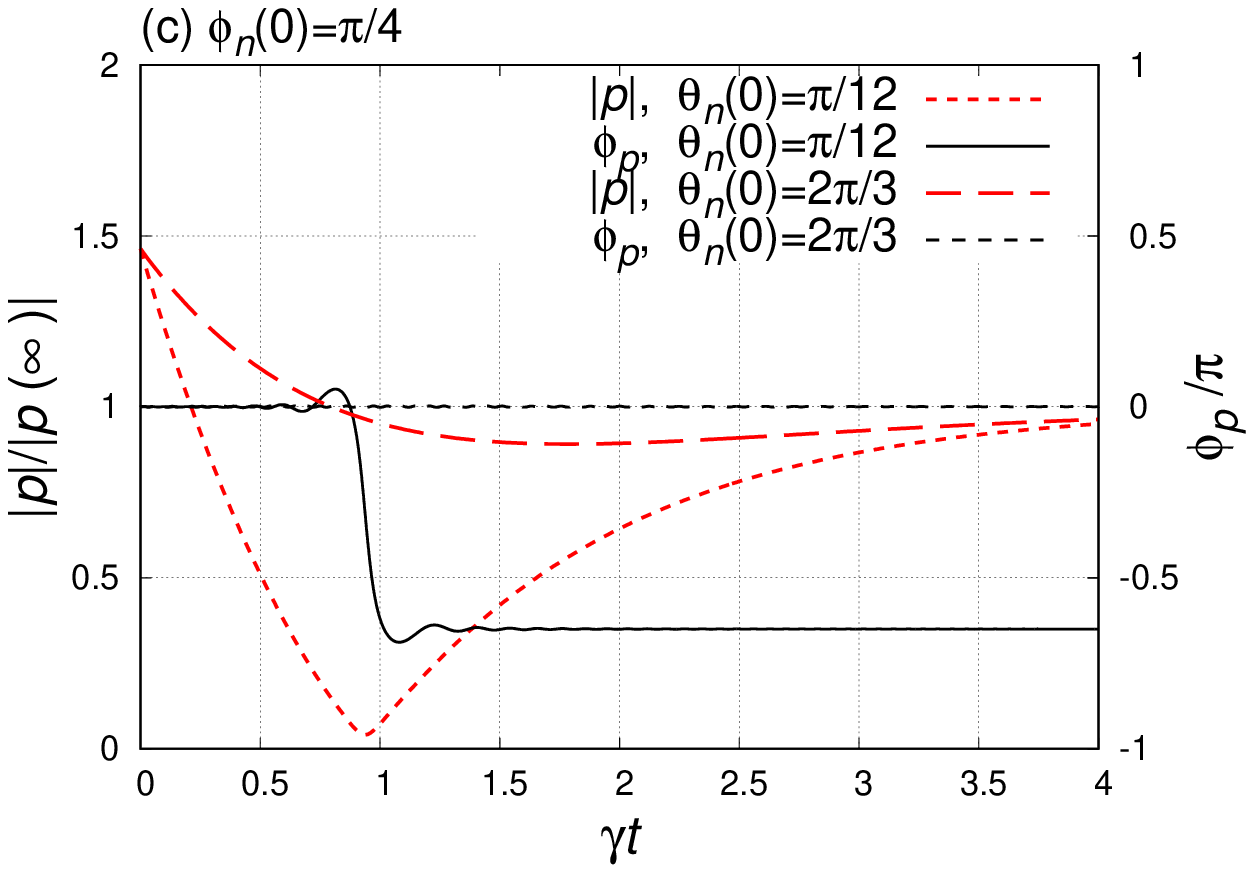} \\
      \includegraphics[width=0.5\linewidth]{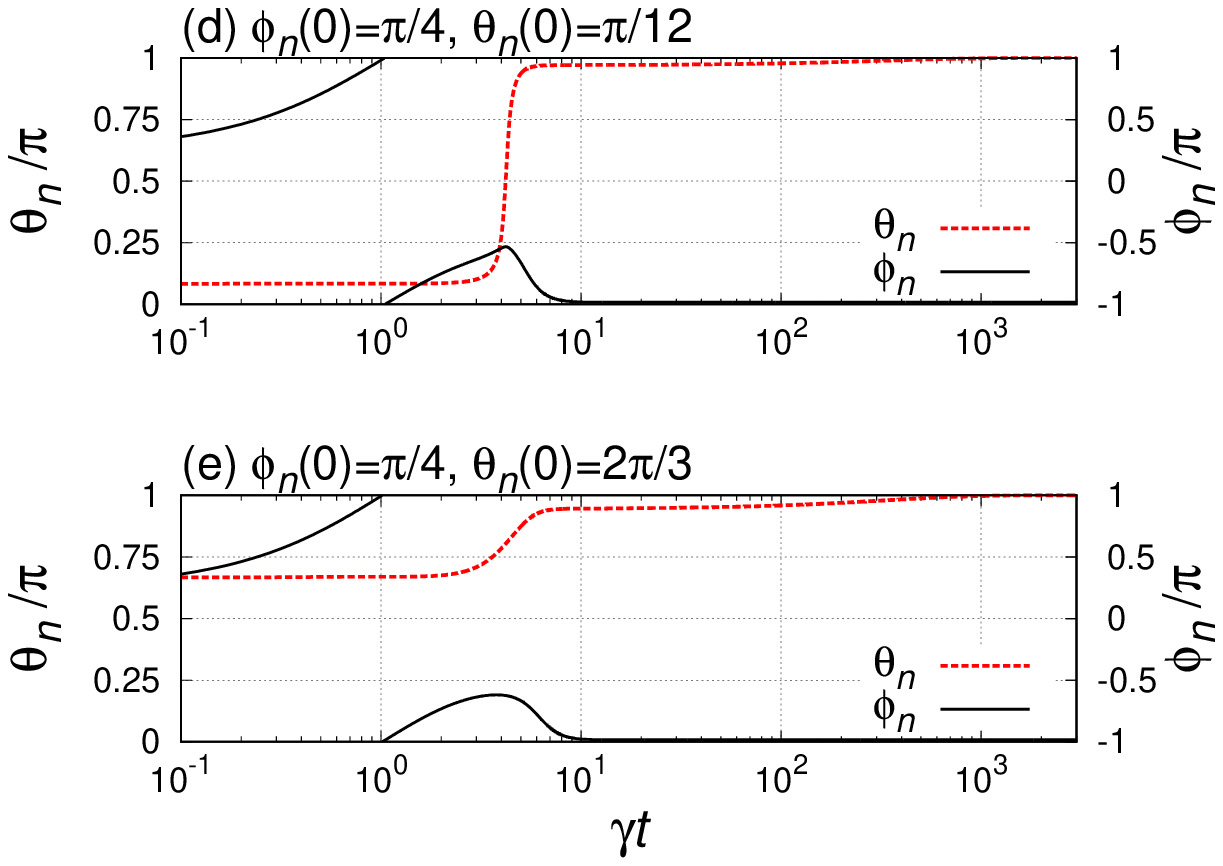} &
      \includegraphics[width=0.5\linewidth]{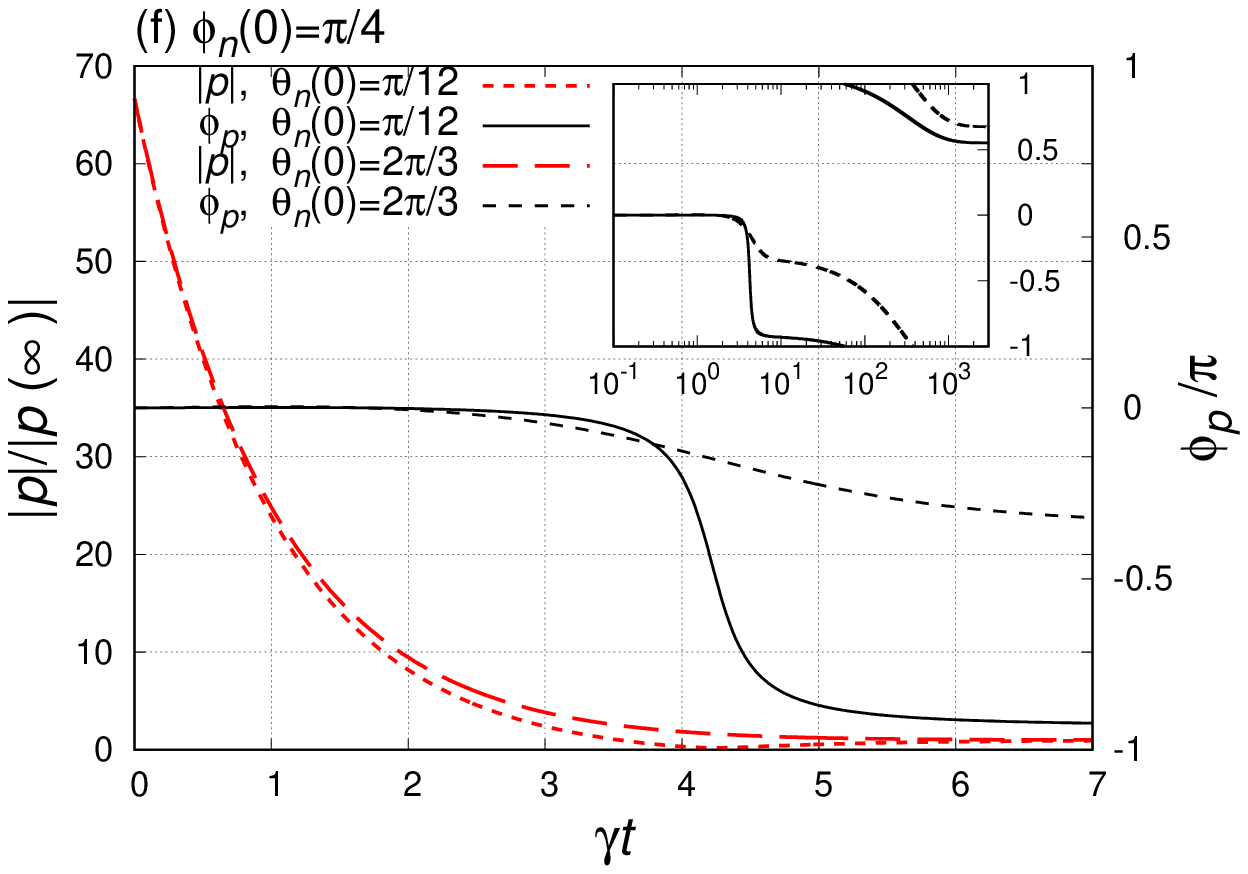}
    \end{tabular}
      \caption{Time evolution of the spin and momentum for two sets of the initial spin angles, namely,
      $\ltk \phi_n(0),\theta_n(0)\rtk=\ltk \pi/4,\pi/12\rtk$ and
      $\ltk \phi_n(0),\theta_n(0)\rtk=\ltk \pi/4,2\pi/3\rtk$, as plotted for
      $\tau_{\rm prec} \ll \tau_{\rm damp}$ ((a), (b), (c)) and $\tau_{\rm prec} \gg \tau_{\rm damp}$ ((d), (e), (f)).
      The definition of each angle is illustrated in Fig.~\ref{fig2}, while the
      parameter values are the same as in Fig.~\ref{fig3}.  For other sets of the initial angles, see appendix~B.}
    \label{fig4}
\end{figure*}

To observe explicitly what happens in between,
we show in Fig.~\ref{fig4} the whole time evolution of the spin and momentum
by taking two sets of the initial spin angles,
$\ltk \phi_n(0),\theta_n(0)\rtk=\ltk \pi/4,\pi/12\rtk$ and $\ltk \pi/4,2\pi/3\rtk$,
which lead to different intermediate behaviors.
For $\tau_{\rm prec} \ll \tau_{\rm damp}$, as can be seen from Figs.~\ref{fig4} (a) and (b),
the spin precesses repeatedly as it should.  Then,
the spin torque is effective at bending the spin direction almost completely to $\theta_n=\pi$
within the damping time scale $t\gamma=\mathcal{O}(1)$.  Simultaneously, as shown in Fig.~\ref{fig4}(c),
the magnitude of $\bp$ reaches $|\bp(\infty)|$ almost completely, which gives the energy minimum of $E_{-1}$.
We observe from Figs.~\ref{fig4}(a) and (c) that
a weird behavior happens when the spin angle $\theta_n$ passes through $\pi/2$:
The precession slows temporarily, while the momentum orientation changes abruptly.
The eventual momentum orientation corresponds to a point just on the ridge that appears in Fig.~\ref{fig3}(a).
This kind of behavior arises presumably from the nonlinearity of the quasi-classical dynamical equations.

For $\tau_{\rm prec}\gg \tau_{\rm damp}$, on the other hand,
we can observe from Figs.~\ref{fig4}(d) and (e) that the spin precesses
only once or less within the damping time scale and that
it takes a long time for the spin orientation to reach $\theta_n=\pi$.
In this case, as shown in Fig.~\ref{fig4}(f), the magnitude of the momentum
decays and almost reaches $|\bp(\infty)|$ within the damping time scale,
whereas its orientation keeps changing continuously even after that in conjunction with the nonlinear spin dynamics.
This is the reason why the asymptotic orientation of the momentum fluctuates
significantly as a function of the initial spin orientations as illustrated
in Fig.~\ref{fig3}(b).
Note also that in the case in which the initial spin is provided along the precession axis,
no spin torque is activated, so that only the momentum relaxes to the point of the energy minimum.
Numerical results for other initial spin angles and also for the time evolution of the velocity are given in appendix~B.

\subsection{Non zero magnetic field: $B\neq 0$}
In the presence of a nonzero magnetic field,
numerical results are qualitatively similar to the $B=0$ case.
There are still minor differences.  For $B\neq0$,
the precession axis deviates from $x$-$y$ plane,
and the precession period $\tau_{\rm prec}$ given by Eq.\ (\ref{prect1}) gets shorter
than the $B=0$ case at a given Rashba coupling strength.
Accordingly, the dynamics of the spin and momentum is modified.
We relegate the numerical results for $B\neq 0$ to appendix~B.



\section{Summary and outlooks}
In the present study we have studied the open-system dynamics of a single spin-$1/2$ particle
with a Rashba-type SOC in a three-dimensional ohmic heat bath
by employing the extended version of the Calderia-Leggett model.
At sufficiently high temperature, we have succeeded in deriving the
quasi-classical Langevin equation for the momentum with a friction term
and the dynamical equation for the instantaneous spin precession;
these equations are nonlinearly coupled with each other, leading to a complex relaxation-dissipation dynamics
until the spin and momentum settle down in one of the minima of the spin-down eigen energy $E_{-1}$.

By obtaining the analytical and numerical solutions to these equations,
we have found that when the precession period (\ref{prect1}) is much shorter
than the damping time (\ref{dampt1}), i.e., $\tau_{\rm prec}\ll \tau_{\rm damp}$,
the initial and final momenta point to almost the same direction
for a wide range of the initial spin direction.
An interesting implication of this finding is that one can control the final state of the spin direction by increasing the Rashba coupling strength $\alpha$.
In the opposite case of $\tau_{\rm prec}\gg \tau_{\rm damp}$, however,
the final momentum significantly fluctuates around the initial one, and so does the final spin direction.
Since the spin dynamics in our model is always accompanied by the momentum dissipation,
the spin relaxation mechanism elucidated in this study is qualitatively different from that
encountered in semiconductors where the Pauli blocking prevents momentum dissipation
and also in nuclear magnetic resonances where the spin is localized.


Throughout the present study we have restricted ourselves to the quasi-classical dynamics that
can be described by deterministic local differential equations by virtue of the ohmic heat bath.
This is a rather ideal case but in general we have to consider memory effects and quantum fluctuations.
For possible application to more realistic cases, e.g., a mobile atomic impurity in trapped cold atoms,
it is not always good to assume high temperatures and ohmic environments,
but dynamical description of the spin and momentum of the impurity may require
the fully quantal time evolution of the reduced density matrix.
For such purpose, the path integral formulation with spin degrees of freedom
that we have developed here may be utilized.
It would be also interesting to use it to explain
a global spin polarization of heavy hadrons observed in relativistic heavy-ion collision experiments \cite{Liang_2005}.

{\it Acknowledgments}---
We would like to thank K.~Nishimura and H.~Yabu for useful discussion.
This work was supported in part by Grants-in-Aid for Scientific Research from JSPS (Nos.\
 17K05445, 18K03501, 18H05406, 18H01211, and 19K14619).

\bibliographystyle{unsrt}
\bibliography{spinorbit_draft_10_17_2021_7_9}

\begin{thebibliography}{10}

\bibitem{landau_lifshitz_nonrelaquantum}
L.~D. Landau and E.~M. Lifshitz.
\newblock {\em QUANTUM MECHANICS , Non-Relativistic Theory 2nd. ed.}
\newblock PERGAMON PRESS, 1965.

\bibitem{On_Closed_Shells}
M.~G. Mayer.
\newblock On closed shells in nuclei.
\newblock {\em Phys. Rev.}, 74:235--239, 1948.

\bibitem{On_the_Magic_Numbers}
O.~Haxel, J.~Hans~D. Jensen, and H.~E. Suess.
\newblock On the "magic numbers" in nuclear structure.
\newblock {\em Phys. Rev.}, 75:1766--1766, 1949.

\bibitem{electro_optic_modulator}
S.~Datta and B.~Das.
\newblock Electronic analog of the electro-optic modulator.
\newblock {\em Appl. Phys. Lett.}, 56:665--667, 1990.

\bibitem{Foldy_Wouthuysen}
Leslie~L. Foldy and Siegfried~A. Wouthuysen.
\newblock On the dirac theory of spin 1/2 particles and its non-relativistic
  limit.
\newblock {\em Phys. Rev.}, 78:29--36, 1950.

\bibitem{Interfacial_charge_spin_coupling}
M.~Johnson and R.~H. Silsbee.
\newblock Interfacial charge-spin coupling: Injection and detection of spin
  magnetization in metals.
\newblock {\em Phys. Rev. Lett.}, 55:1790--1793, 1985.

\bibitem{sinova}
J.~Sinova, D.~Culcer, Q.~Niu, N.~A. Sinitsyn, T.~Jungwirth, and A.~H.
  MacDonald.
\newblock Universal intrinsic spin hall effect.
\newblock {\em Phys. Rev. Lett.}, 92:126603, 2004.

\bibitem{rashba_semiconductor}
Yu.~A. Bychkov, V.~I. Mel'nikovand, and E.~I. Rashba.
\newblock Effect of spin-orbit coupling on the energy spectrum of a 2d electron
  system in a tilted magnetic field.
\newblock {\em Zh. Eksp. Teor. Fiz.}, 98:717--726, 1990.

\bibitem{rashba_socoupling}
Yu.~A. Bychkov and E.~I. Rashba.
\newblock Properties of a electron gas with lifted spectral degeneracy.
\newblock {\em Zh. Eksp. Teor. Fiz.}, 39:66--69, 1984.

\bibitem{dresselhaus}
G.~Dresselhaus.
\newblock Spin-orbit coupling effects in zinc blende structures.
\newblock {\em Phys. Rev.}, 100:580--586, 1955.

\bibitem{so-coupling_zeemaneffect}
Tomohiro Yokoyama, Mikio Eto, and Yuli~V. Nazarov.
\newblock Anomalous josephson effect induced by spin-orbit interaction and
  zeeman effect in semiconductor nanowires.
\newblock {\em Phys. Rev. B}, 89:195407, 2014.

\bibitem{so-coupling_rotationaleffect}
Konstantin~Y. Bliokh, Yuri Gorodetski, Vladimir Kleiner, and Erez Hasman.
\newblock Coriolis effect in optics: Unified geometric phase and spin-hall
  effect.
\newblock {\em Phys. Rev. Lett.}, 101:030404, Jul 2008.

\bibitem{Stanescu_Galitski_2007}
T.~D. Stanescu, C.~Zhang, and V.~Galitski.
\newblock Nonequilibrium spin dynamics in a trapped fermi gas with effective
  spin-orbit interactions.
\newblock {\em Phys. Rev. Lett.}, 99(11):110403, 2007.

\bibitem{SOBEC}
Y.~J. Lin, K.~Jiménez-García, and I.~B. Spielman.
\newblock Spin-orbit-coupled bose-einstein condensates.
\newblock {\em Nature}, 471:83–86, 2011.

\bibitem{PhysRevLett.102.230402}
Andr\'e Schirotzek, Cheng-Hsun Wu, Ariel Sommer, and Martin~W. Zwierlein.
\newblock Observation of fermi polarons in a tunable fermi liquid of ultracold
  atoms.
\newblock {\em Phys. Rev. Lett.}, 102:230402, Jun 2009.

\bibitem{PhysRevLett.117.055301}
Ming-Guang Hu, Michael~J. Van~de Graaff, Dhruv Kedar, John~P. Corson, Eric~A.
  Cornell, and Deborah~S. Jin.
\newblock Bose polarons in the strongly interacting regime.
\newblock {\em Phys. Rev. Lett.}, 117:055301, Jul 2016.

\bibitem{PhysRevLett.117.055302}
Nils~B. J\o{}rgensen, Lars Wacker, Kristoffer~T. Skalmstang, Meera~M. Parish,
  Jesper Levinsen, Rasmus~S. Christensen, Georg~M. Bruun, and Jan~J. Arlt.
\newblock Observation of attractive and repulsive polarons in a bose-einstein
  condensate.
\newblock {\em Phys. Rev. Lett.}, 117:055302, Jul 2016.

\bibitem{Lampo}
A.~Lampo, S.~H. Lim, M.~Á. García-March, and M.~Lewenstein.
\newblock Bose polaron as an instance of quantum brownian motion.
\newblock {\em Quantum}, 1:30, 2017.

\bibitem{Boyanovsky_2019}
D.~Boyanovsky, D.~Jasnow, X.~Lun Wu, and R.~C. Coalson.
\newblock Dynamics of relaxation and dressing of a quenched bose polaron.
\newblock {\em Phy. Rev. A}, 100:043617, 2019.

\bibitem{2019NJPh...21d3014K}
K.~{Knakkergaard Nielsen}, L.~A. {Pe{\~n}a Ardila}, G.~M. {Bruun}, and
  T.~{Pohl}.
\newblock {Critical slowdown of non-equilibrium polaron dynamics}.
\newblock {\em New Journal of Physics}, 21(4):043014, April 2019.

\bibitem{Dyakonov1971}
M.~Dyakonov and V.~Perel.
\newblock Spin orientation of electrons associated with the interband
  absorption of light in semiconductors.
\newblock {\em SOV. PHYS. JETP}, 33:1053, 1971.

\bibitem{Bloembergen_NMR}
N.~Bloembergen, E.~M. Purcell, and R.~V. Pound.
\newblock Relaxation effects in nuclear magnetic resonance absorption.
\newblock {\em Phys. Rev.}, 73:679--712, 1948.

\bibitem{kubo_tomita_NMR}
Ryogo Kubo and Kazuhisa Tomita.
\newblock A general theory of magnetic resonance absorption.
\newblock {\em J. Phys. Soc. Japan}, 9(6):888--919, 1954.

\bibitem{Solomon_spinrelaxation}
I.~Solomon.
\newblock Relaxation processes in a system of two spins.
\newblock {\em Phys. Rev.}, 99:559--565, 1955.

\bibitem{Caldeira1983}
A.~O. Caldeira and A.~J. Leggett.
\newblock Path integral approach to quantum brownian motion.
\newblock {\em Physica A}, 121(3):587--616, 1983.

\bibitem{Ambegaokar}
V.~Ambegaokar.
\newblock Quantum brownian motion and its classical limit.
\newblock {\em Ber. Bunsenges. Phys. Chem.}, 95(3):400--404, 1991.

\bibitem{DIOSI1993517}
L.~Diósi.
\newblock Calderia-leggett master equation and medium temperatures.
\newblock {\em Physica A}, 199(3):517--526, 1993.

\bibitem{Schlosshauer}
M.~A. Schlosshauer.
\newblock {\em Decoherence and the Quantum-To-Classical Transition}.
\newblock Springer, 2007.

\bibitem{schmid}
A.~Schmid.
\newblock On a quasiclassical langevin equation.
\newblock {\em J. Low Temp. Phys.}, 49:609, 1982.

\bibitem{lindblad}
G.~Lindblad.
\newblock On the generators of quantum dynamical semigroups.
\newblock {\em Commun. in Math. Phys.}, 48:119, 1976.

\bibitem{Gao_PhysRevLett.79.3101}
S.~Gao.
\newblock Dissipative quantum dynamics with a lindblad functional.
\newblock {\em Phys. Rev. Lett.}, 79:3101, 1997.

\bibitem{Feynman}
R.~P. Feynman and F.~L. Vernon.
\newblock The theory of a general quantum system interacting with a linear
  dissipative system.
\newblock {\em Ann. Phys.}, 24:118--173, 1963.

\bibitem{altland_simons}
A.~Altland and B.~Simons.
\newblock {\em Condensed Matter Field Theory 2nd ed.}
\newblock CAMBRIDGE UNIVERSITY PRESS, 2010.

\bibitem{Zurek}
W.~H. Zurek.
\newblock Reduction of the wavepacket: How long does it take ?
\newblock {\em Los Alamos report LAUR 84-2750}, first pulished in 1984.

\bibitem{RevModPhys_zurek}
W.~H. Zurek.
\newblock Decoherence, einselection, and the quantum origins of the classical.
\newblock {\em Rev. Mod. Phys.}, 75:715--775, 2003.

\bibitem{Liang_2005}
Zuo-Tang Liang and Xin-Nian Wang.
\newblock Globally polarized quark-gluon plasma in noncentrala+acollisions.
\newblock {\em Physical Review Letters}, 94(10), Mar 2005.

\end{thebibliography}

\newpage
\appendix
\section{Off-diagonal fluctuation effects}
We rewrite the effective action (\ref{effact1}) for the ohmic case in terms of $\bx_\pm$ and $g_\pm$ as
\begin{eqnarray}
  && iW\ldk \bx_+ g_+, \bx_- g_-\rdk
 \nn &=&
   i\int^t_0 {\rm d}u\, \ltk
   m\lk \dot{\bx}_+ - d\bn_+\times \bm{\alpha} \rk
   \cdot
   \lk \dot{\bx}_- - d\bn_- \times \bm{\alpha} \rk
   \right.
   \nn
&&\quad \qquad -\dot{\phi}_+\sin \theta_+ \sin\frac{\theta_-}{2}
      +\frac{1}{2}\dot{\phi}_-\cos \theta_+ \cos\frac{\theta_-}{2}
\nn
&&\quad \qquad -d\bn_-\cdot \bm{B} +\, c\,  \left. \dot{\bx}_+ \cdot \bx_-\rtk
\nn
&&
 -\frac{1}{2} \int^t_0 {\rm d}u\,  \int^t_0 {\rm d}u' \,   L(u-u') \bx_-(u)\cdot \bx_-(u'),
  \label{effact2}
\end{eqnarray}
where $d\bn_+=\ltk \bn(g) + \bn(g')\rtk/2$, $d\bn_-=\bn(g) -\bn(g')$, and
\beq
L(u-u')=\int^{\infty}_0 {\rm d}\omega \,
J(\omega)\, \coth{\frac{\omega \beta}{2} }
\, \cos{\omega (u-u')}.
\label{actapp1}
\eeq
Since we are interested in the probability functional of the quasi-classical path $\bx_+$,
we integrate out the off-diagonal fluctuation path $\bx_-$, which is suppressed by the Gaussian term
including $L(u-u')$, to obtain the probability functional up to an irrelevant constant as
\beq
&&i\tilde{W}\ldk \bx_+g_+, g_-\rdk
\nn
&=&
   -i\int^t_0 {\rm d}u\, \ltk
   m\lk \dot{\bx}_+ -d\bn_+\times \bm{\alpha} \rk \cdot \lk  d\bn_- \times \bm{\alpha} \rk
   \right.
   \nn
&&  \quad + \dot{\phi}_+\sin \theta_+ \sin\frac{\theta_-}{2}
      -\frac{1}{2}\dot{\phi}_-\cos \theta_+ \cos\frac{\theta_-}{2}
\nn
&& \left. \quad\quad+d\bn_-\cdot \bm{B} \rtk
\nn
&& +i \ltk m(\dot{\bx}_+(t) - d \bn_+(t)\times\bm{\alpha})\cdot\bx_-(t) \right.
\nn
 && \quad \left. - m(\dot{\bx}_+(0) - d \bn_+(0)\times\bm{\alpha})\cdot\bx_-(0) \rtk
\nn
&&
   - \frac{1}{2} \int^t_0 {\rm d}u\,  \int^t_0 {\rm d}u' \,   L^{-1}(u-u')
   \nn
 &&\quad \times \bm{\xi}\ldk \bx_+(u),d\bn_+(u)\rdk \cdot \bm{\xi}\ldk \bx_+(u'),d\bn_+(u')\rdk,
  \label{effact4}
\eeq
where we have introduced a {\it noise} functional
\beq
\bm{\xi}\ldk \bx_+, d\dot{\bn}_+\rdk \equiv m\ldk \ddot{\bx}_+
+\gamma \dot{\bx}_+ - d\dot{\bn}_+\times \bm{\alpha} \rdk.
\label{noise1}
\eeq
The above result can be interpreted as follows:
If the functional $\bm{\xi}$ is used instead of $\bx_+$
in the path integral of $e^{i\tilde{W}}$, i.e.,
${\mathcal D}\bx_+ = {\mathcal J D}\bm{\xi}$
with ${\mathcal J}=|{\mathcal D}\bx_+/{\mathcal D}\bm{\xi}|$ the functional Jacobian,
$\bx_+$ becomes inversely the functional of $\bm{\xi}$
whose path probability is given by ${\mathcal J}\, e^{i\tilde{W}}$.
The Jacobian in our case becomes a constant.
Once a noise fluctuation path $\bm{\xi}$ is given in accordance with this probability,
therefore, Eq.\ (\ref{noise1}) can be regarded as a Langevin equation, whose solution
for $\dot{\bx}_+$ is given by
\beq
\dot{\bx}_+(u) &=& \dot{\bx}_+(0) e^{-\gamma u}
\nn
&&+\int_0^{u} {\rm d}u'\,
e^{-\gamma (u-u')}
\ldk \frac{\bm{\xi}(u')}{m}
+d\dot{\bn}_+(u')\times \bm{\alpha}\rdk
\nn
&=& d\bn_+(u)\times \bm{\alpha}
+e^{-\gamma u} \ldk \dot{\bx}_+(0) -d\bn_+(0)\times \bm{\alpha}\rdk
\nn
&&-\gamma\, \int_0^{u} {\rm d}u'\,
e^{-\gamma (u-u')} d\bn_+(u')\times \bm{\alpha}
\nn
&&+\frac{1}{m} \int_0^{u} {\rm d}u'\,
e^{-\gamma (u-u')}
\bm{\xi}(u').
\label{invfn1}
\eeq
Plugging the above expression into Eq.~(\ref{effact4}),
we obtain the functional in terms of $\bm{\xi}$ and $g_\pm$ as
\begin{eqnarray}
  && i\tilde{W}\ldk \bm{\xi}; g_\pm \rdk
  =
     -i\int^t_0 {\rm d}u\, \ltk
     e^{-\gamma u} \bp_+(0) \cdot \lk  d\bn_- \times \bm{\alpha} \rk
     \right.
  \nn
  && +\dot{\phi}_+\sin \theta_+ \sin\frac{\theta_-}{2}
        -\frac{1}{2}\dot{\phi}_-\cos \theta_+ \cos\frac{\theta_-}{2}
        +d\bn_-\cdot \left.  \bm{B} \rtk
  \nn
  &&
  +i\gamma m \, \int^t_0 {\rm d}u\, \int_0^{t} {\rm d}u'\,
  G(u-u')
  \nn
  &&
  \qquad \qquad\qquad \ \times
  \ltk d\bn_+(u')\times \bm{\alpha}\rtk \cdot \ltk  d\bn_-(u) \times \bm{\alpha} \rtk
  \nn
  &&
   -i\, \int^t_0 {\rm d}u\, \int_0^{t} {\rm d}u'\,
   \bm{\xi}(u') \cdot \ltk  G(u-u') d\bn_-(u) \times \bm{\alpha} \right.
   \nn
   &&
   \quad\qquad \qquad  \qquad \qquad - \left. \delta (u-u') G(t-u') \bx_-(t) \rtk
  \nn
  && -i m\gamma \int^t_0 {\rm d}u \, G(t-u) \ltk {\rm d}\bn_+(u)\times\bm{\alpha}\rtk \cdot\bx_-(t)
   \nn
  && - i\bm{p}_+(0) \cdot \bx_-(0)+ i e^{-\gamma t} \bm{p}_+(0) \cdot\bx_-(t)
   \nn
  &&
  \ -\frac{1}{2} \int^t_0 {\rm d}u\,  \int^t_0 {\rm d}u' \,   L^{-1}(u-u') \, \bm{\xi}(u) \cdot \bm{\xi}(u'),
\label{effact5}
\end{eqnarray}
where $G(t)=e^{-\gamma t}\theta(t)$ is the retarded Green function.
Note that the path $\bm{\xi}=0$ corresponds to the quasi-classical path of $\bx_+$
that satisfies the homogeneous equation of Eq.~(\ref{noise1}).
From the Gaussian form for $\bm{\xi}$ in Eq.~(\ref{effact4})
we expect this quasi-classical path to dominate the path integral
when the off-diagonal Euler-angle fluctuations are consistently small, i.e., $g_-\sim 0$.
In this case the noise fluctuation satisfies $\langle \bm{\xi} \rangle=0$
and a Brownian (Kubo's second) fluctuation-dissipation relation
$\langle \xi_i(t) \xi_j(t')\rangle = \delta_{ij}L(t-t') $.

Finally, performing the Gaussian path integral with respect to $\bm{\xi}$,
we obtain the effective action for the spin variables $g_\pm$ as
\beq
&&i\mathcal{W}\ldk g_\pm \rdk
=
   -i\int^t_0 {\rm d}u\,
 \ltk \dot{\phi}_+\sin \theta_+ \sin\frac{\theta_-}{2} \right.
 \nn
 && \left.
      -\frac{1}{2}\dot{\phi}_-\cos \theta_+ \cos\frac{\theta_-}{2}
      +d\bn_-\cdot \bm{B} \rtk
\nn
&&
-i \bp_+(0) \cdot \ltk \bm{\beta}_-(0) - e^{-\gamma t} \bx_-(t)+\bx_-(0) \rtk
\nn
&&
+i\gamma m \, \int_0^{t} {\rm d}u\,
\ltk d\bn_+(u)\times \bm{\alpha}\rtk \cdot \ltk \bm{\beta}_-(u) - G(t-u) \bx_-(t) \rtk
\nn
&&
-\frac{1}{2}
\int_0^{t} {\rm d}u \int_0^{t} {\rm d}u'\, L(u-u')
\nn
 &&
 \times
 \ltk \bm{\beta}_-(u) - G(t-u) \bx_-(t) \rtk \cdot \ltk \bm{\beta}_-(u') - G(t-u') \bx_-(t) \rtk,
 \nn
\label{effact6}
\eeq
where we have introduced the following functional
\beq
\bm{\beta}_-(u) = \int^t_0 {\rm d}u'\, G(u'-u)  \ltk  d\bn_-(u') \times \bm{\alpha} \rtk.
\eeq
In Eq.~(\ref{effact6}) we again encounter the Gaussian suppression
characterized by $L(u-u')$ also for the off-diagonal spin variable $g_-$.
At high temperatures where $L(u-u')\simeq 2\gamma m k_B T \delta(u-u')$,
this Gaussian term
can be approximated by
\beq
 &&-\frac{1}{2}\int_0^{t} {\rm d}u \int_0^{t} {\rm d}u'\, L(u-u')
 \nn
 &&
\times
  \ltk \bm{\beta}_-(u) - G(t-u) \bx_-(t) \rtk
  \cdot \ltk \bm{\beta}_-(u') - G(t-u') \bx_-(t) \rtk
  \nn
&&
\simeq
-m \gamma k_BT
\int_0^{t} {\rm d}s\, \int_0^{t} {\rm d}s'\, \int_0^{t} {\rm d}u\,
G(s-u)G(s'-u)
   \nn
   &&
   \quad\times \ltk d\bm{n}_-(s)\times\bm{\alpha} - 2\delta (t-s) \bm{x}_-(s) \rtk
   \nn
   &&
   \quad\cdot \ltk d\bm{n}_-(s')\times\bm{\alpha} - 2\delta (t-s') \bm{x}_-(s') \rtk
   \nn
   &&
   =-2m k_BT \int_0^{t} {\rm d}s\, \int_0^{s} {\rm d}s'\,
   e^{-\gamma s} \sinh\lk \gamma s'\rk
   \nn
    &&
    \quad \times\ltk d\bn_-(s)\times\bm{\alpha} - 2\delta(t-s) \bx_-(s) \rtk
    \nn
    &&
    \quad\cdot\ltk d\bn_-(s')\times\bm{\alpha} - 2\delta(t-s') \bx_-(s') \rtk.
 \label{gmgauss1}
\eeq
Indeed the above expression is of Gaussian form, but it is nonlocal in time unlike the case of $\bx_-$ as depicted in Eq.~(\ref{w1hta1}).
This nonlocality prevents us from explicitly proving the decoherence of the off-diagonal spin variable $g_-$.
Nevertheless, one can show that after integrating out the orbital variables $\bx_\pm$, the path of $g_-=0$, i.e., $\beta_-=0$, is not
suppressed by any Gaussian factor, implying that it gives the most probable one, i.e., the quasi-classical path.
In fact, the variation of ${\mathcal W}$ with respect to $g_-$ at $g_-=0$ restores Eq.~(\ref{dotn_beff}) that has Eq.~(\ref{beffsol1})
incorporated.  We postpone further detailed analysis about nonlinear
fluctuation effects of $g_-$ on the quasi-classical dynamics
elsewhere.

\section{Time evolution of spin and momentum}
\begin{figure*}[!htbp]
  \centering
    \begin{tabular}{rr}
      \vspace{0.1cm} \\
      \includegraphics[width=0.5\linewidth,height=5.5cm]{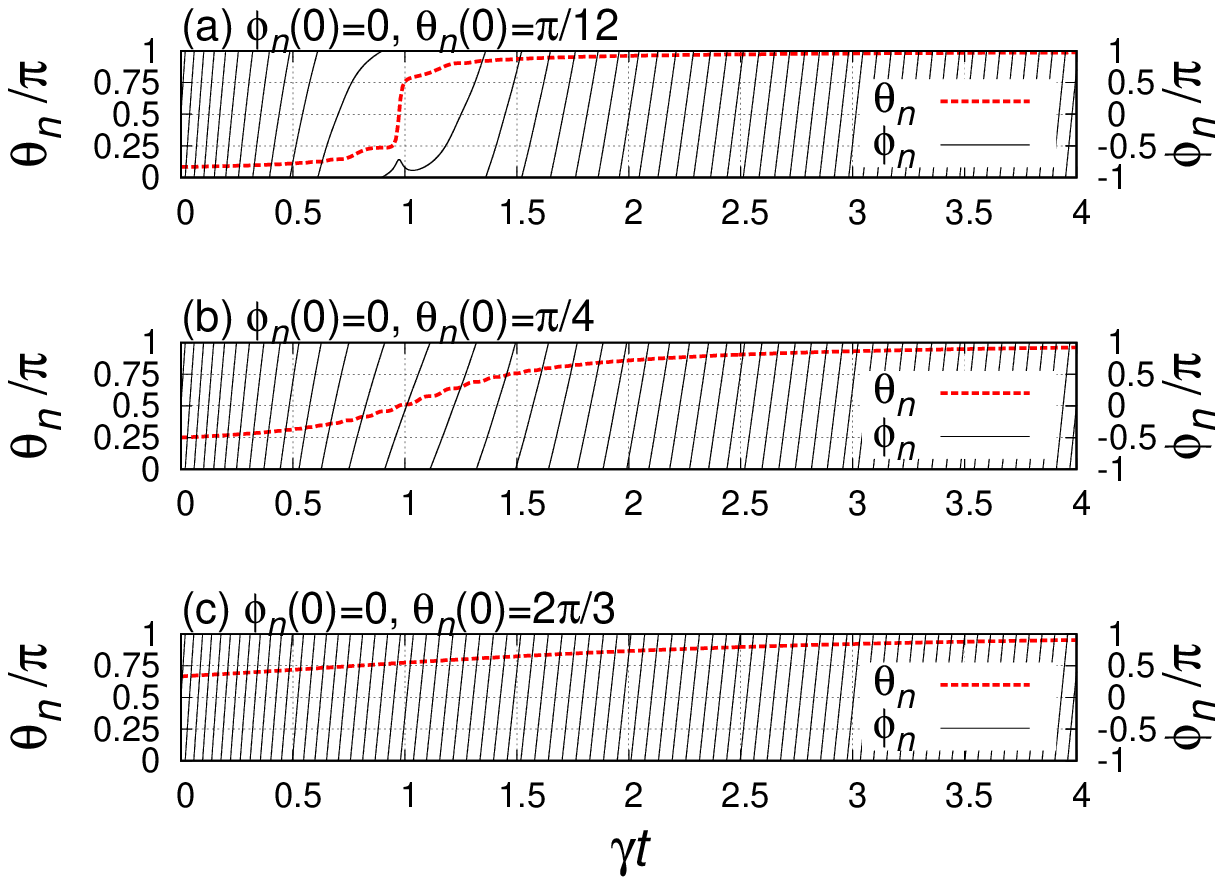} &
      \includegraphics[width=0.5\linewidth,height=5.5cm]{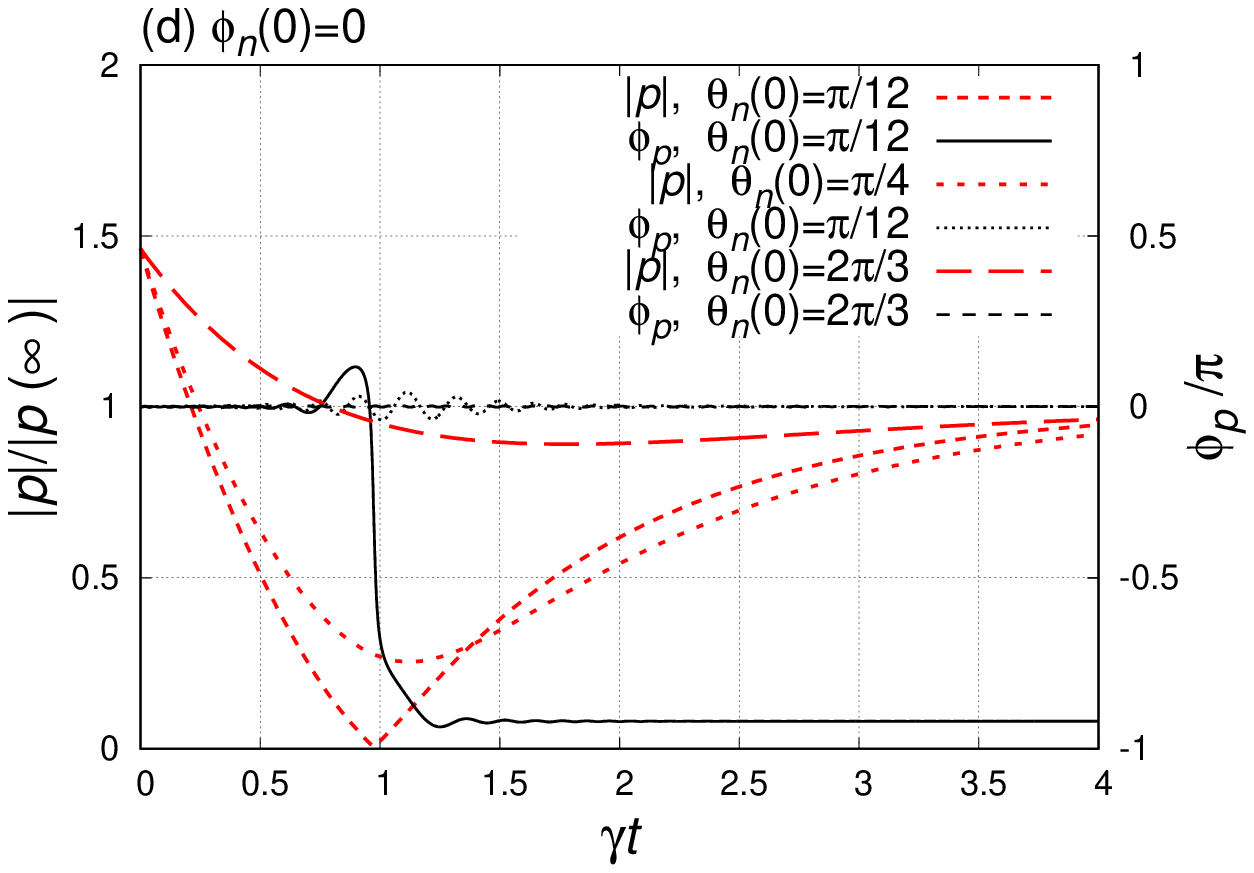} \\
      \vspace{0.1cm}\quad \\
      \includegraphics[width=0.5\linewidth,height=5.5cm]{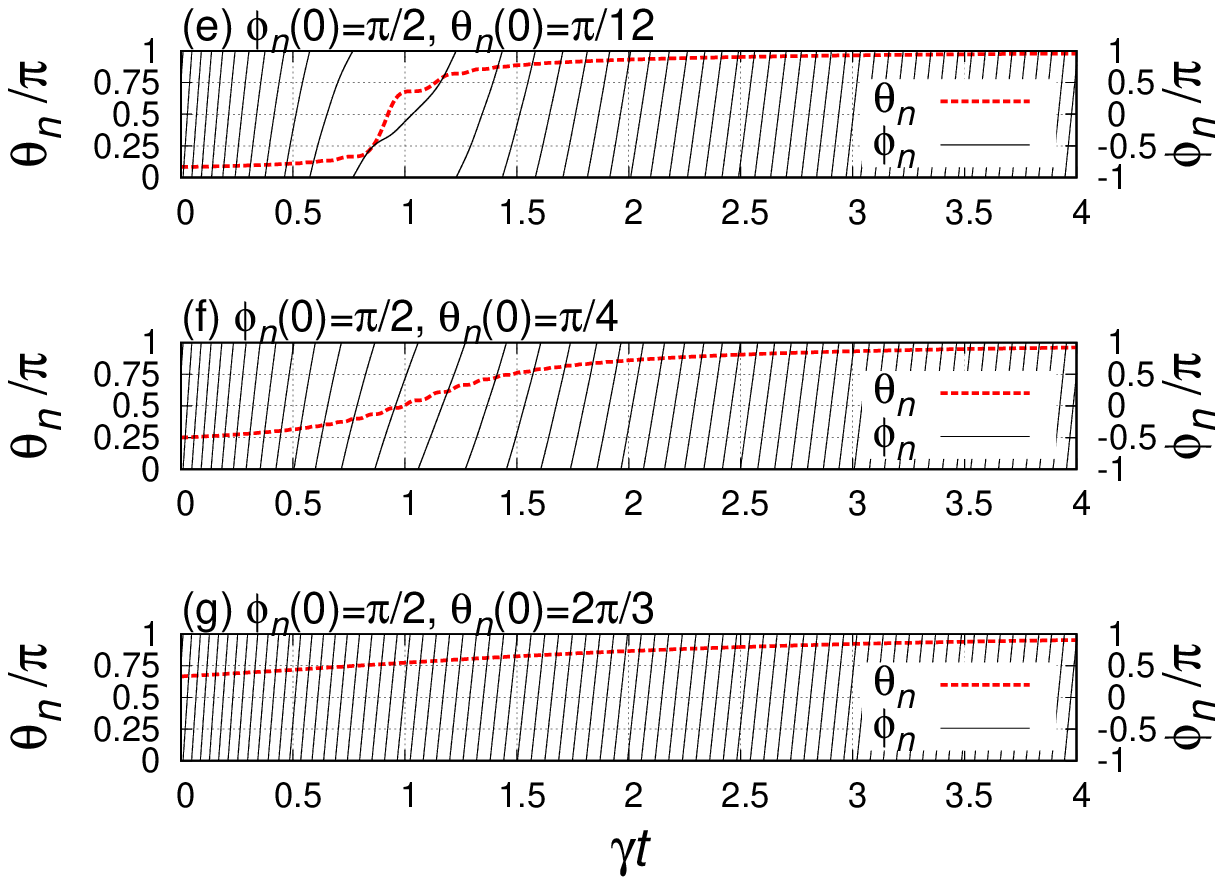} &
      \includegraphics[width=0.5\linewidth,height=5.5cm]{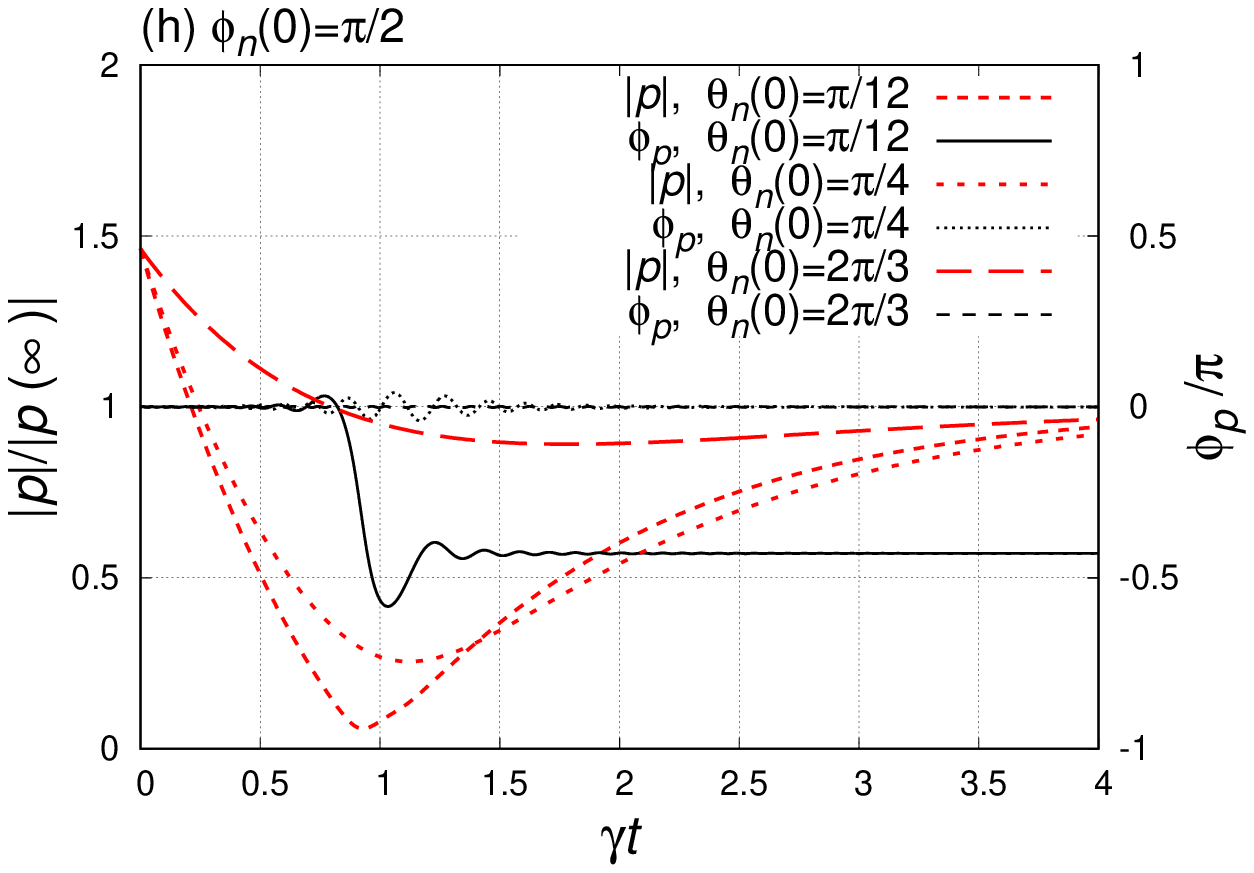} \\
    \end{tabular}
      \caption{Time evolution of the spin and momentum at $B=0$, as plotted for $\tau_{\rm prec} \ll \tau_{\rm damp}$
      under the initial conditions
      $\ltk \phi_n(0), \theta_n(0) \rtk =\ltk 0,\ltk \pi/12,\pi/4,2\pi/3 \rtk \rtk $ ((a), (b), (c), (d))
       and $\ltk \phi_n(0), \theta_n(0) \rtk =\ltk \pi/2,\ltk \pi/12,\pi/4,2\pi/3 \rtk \rtk $ ((e), (f), (g), (h)).}
    \label{fig6}
\end{figure*}

\begin{figure*}[htbp]
  \centering
    \begin{tabular}{rr}
      \includegraphics[width=0.5\linewidth,height=5.5cm]{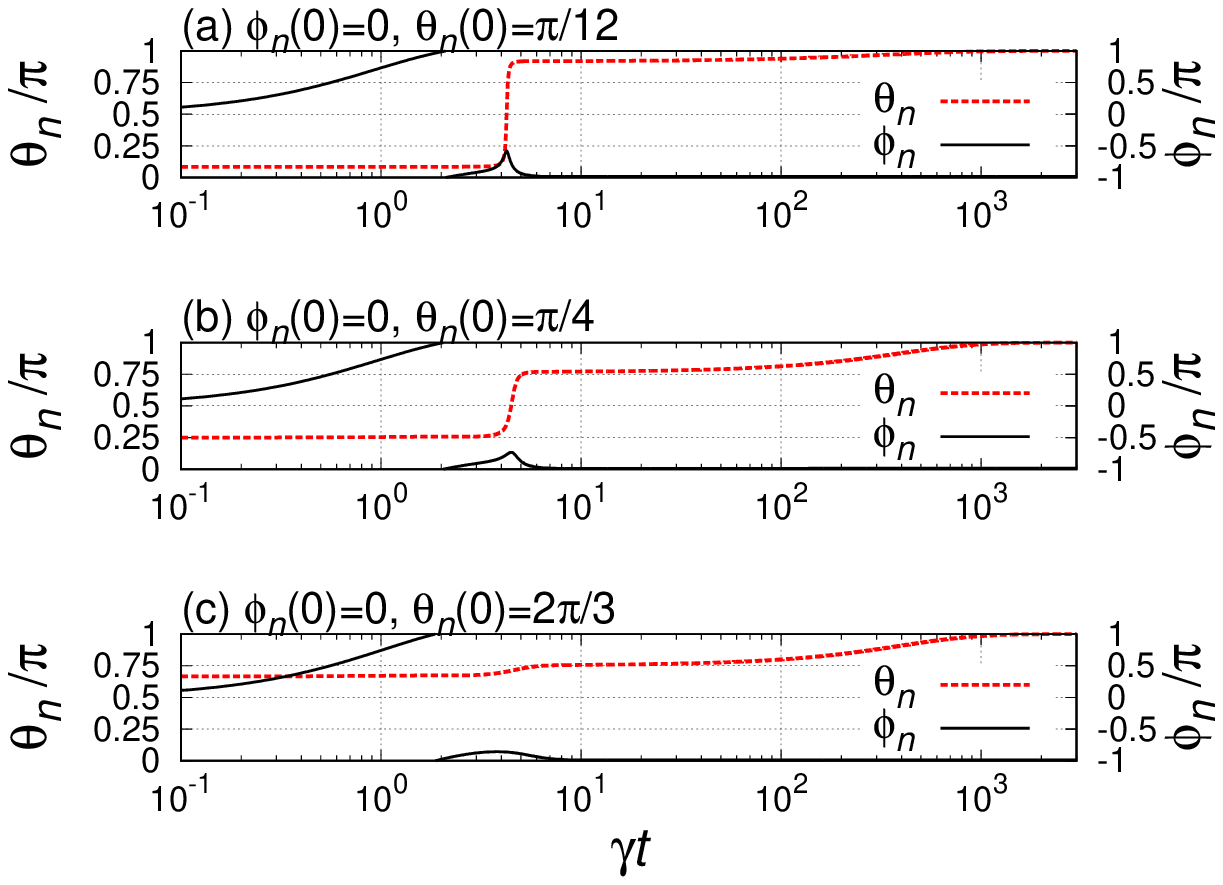} &
      \includegraphics[width=0.5\linewidth,height=5.5cm]{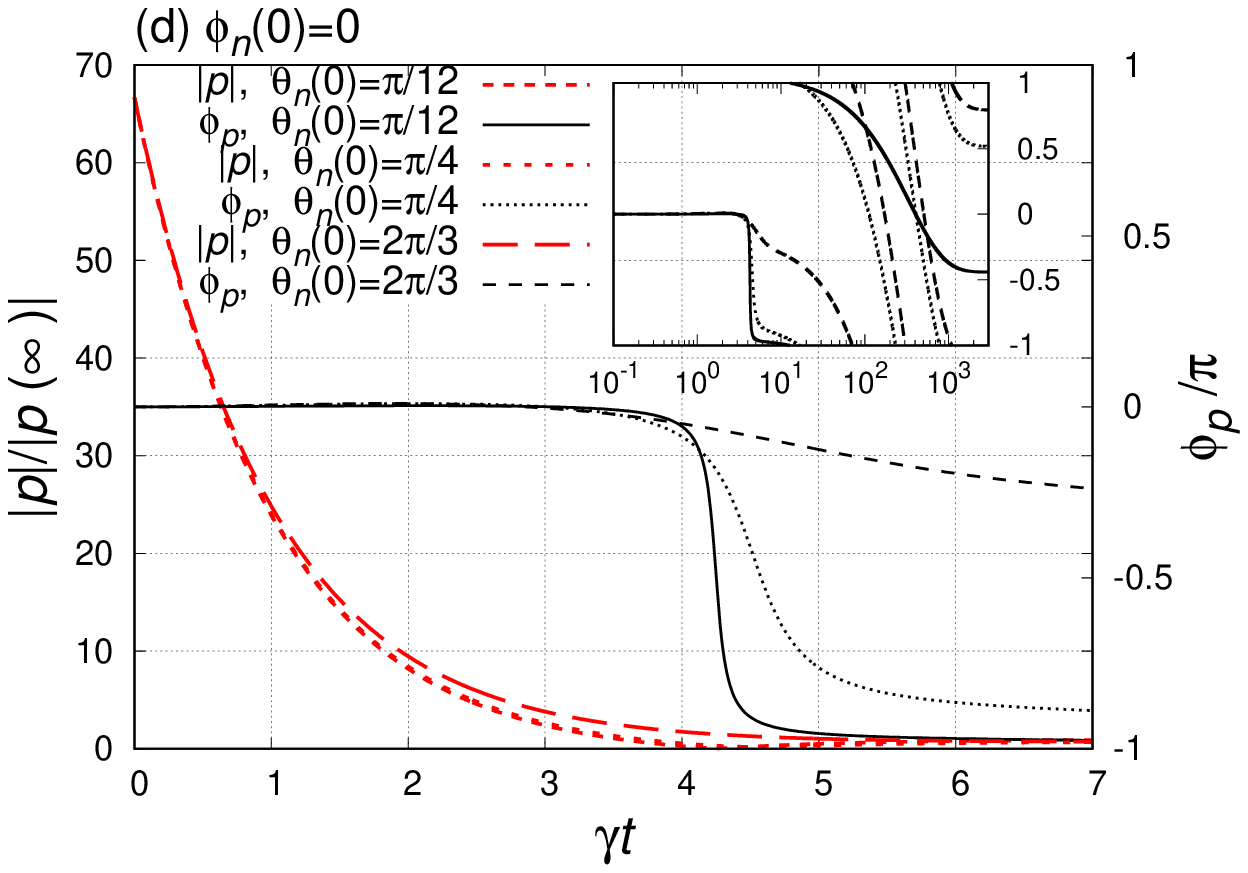} \\
      \vspace{0.1cm}\quad \\
      \includegraphics[width=0.5\linewidth,height=5.5cm]{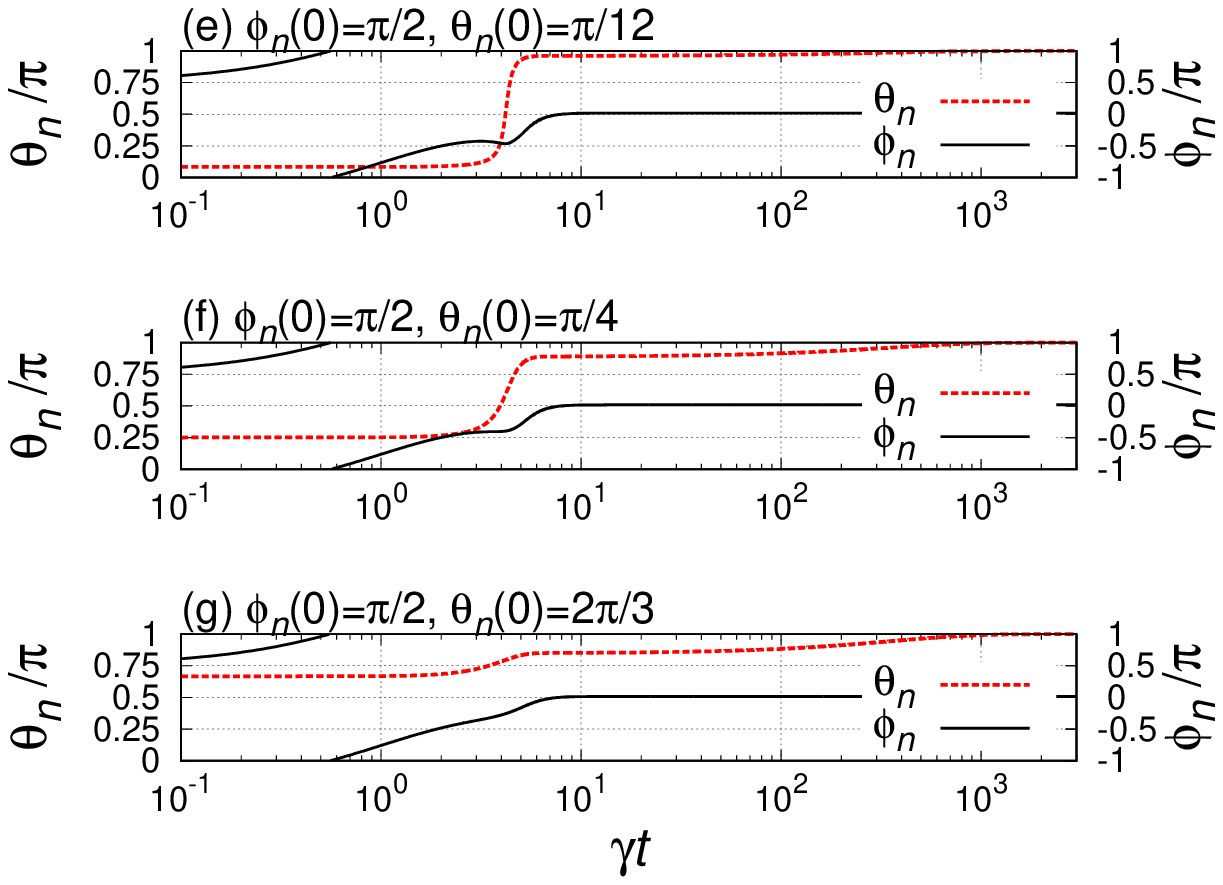} &
      \includegraphics[width=0.5\linewidth,height=5.5cm]{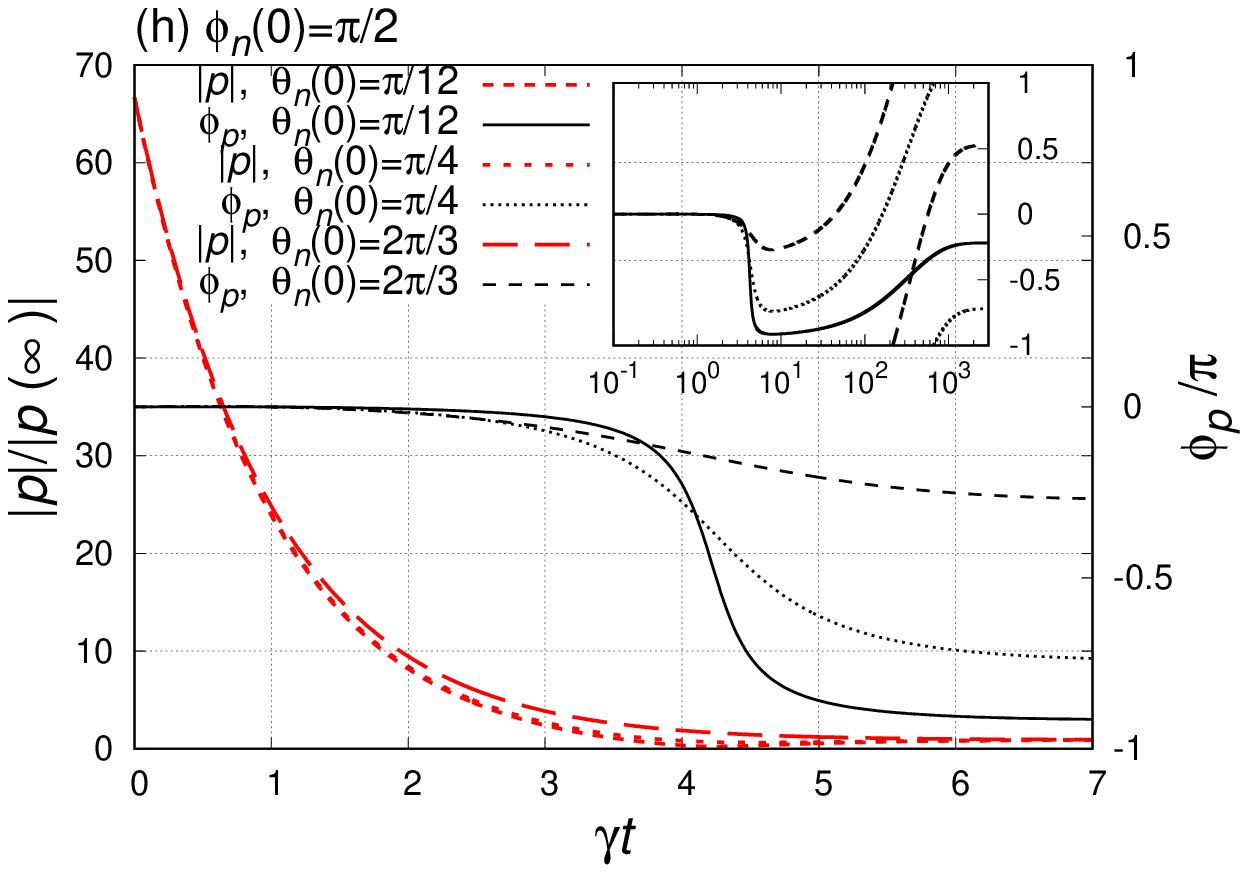} \\
    \end{tabular}
      \caption{Time evolution of the spin and momentum at $B=0$, as plotted for $\tau_{\rm prec} \gg \tau_{\rm damp}$
      under the initial conditions
      $\ltk \phi_n(0), \theta_n(0) \rtk =\ltk 0,\ltk \pi/12,\pi/4,2\pi/3 \rtk \rtk $ ((a), (b), (c), (d))
      and $\ltk \phi_n(0), \theta_n(0) \rtk =\ltk \pi/2,\ltk \pi/12,\pi/4,2\pi/3 \rtk \rtk $ ((e), (f), (g), (h)).}
    \label{fig7}
\end{figure*}

\begin{figure*}[htbp]
    \centering
    \begin{tabular}{rr}
      \includegraphics[width=0.5\linewidth]{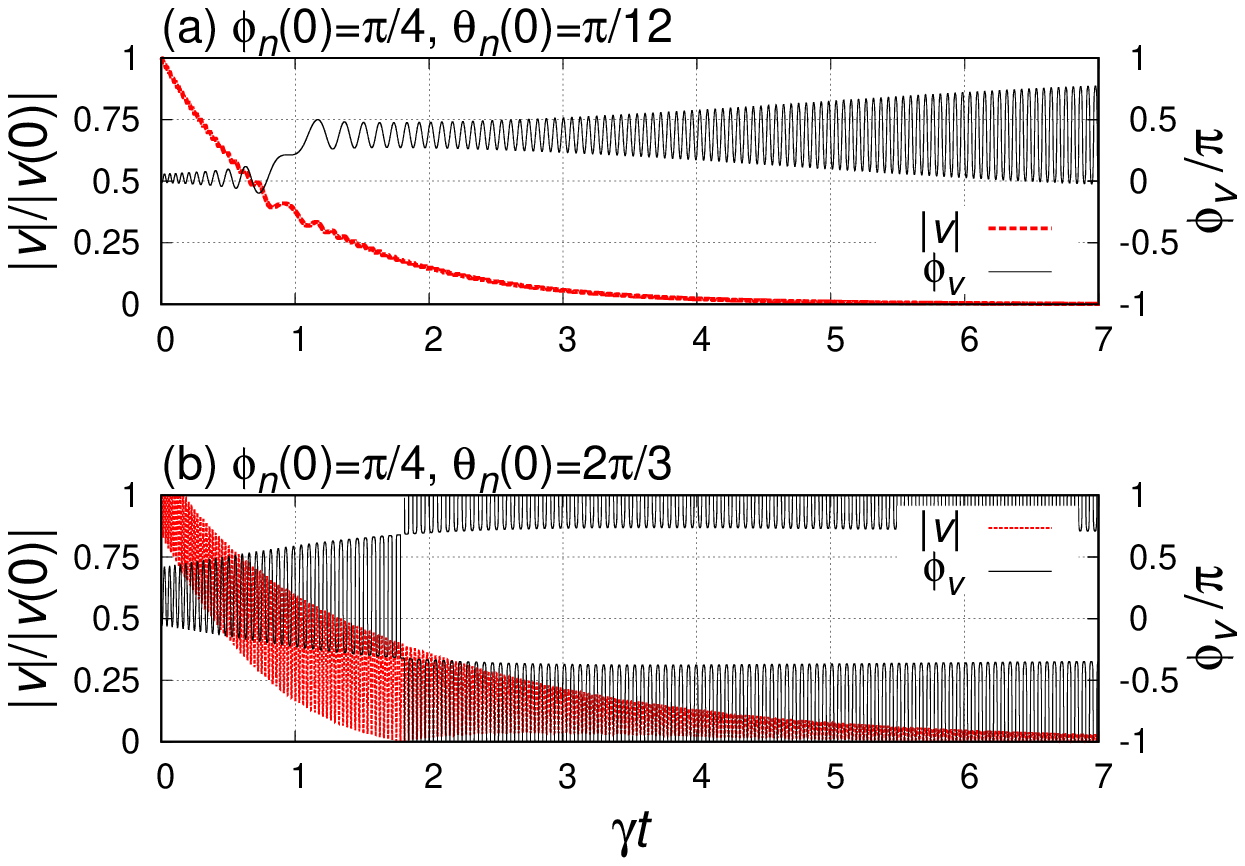} &
      \includegraphics[width=0.5\linewidth]{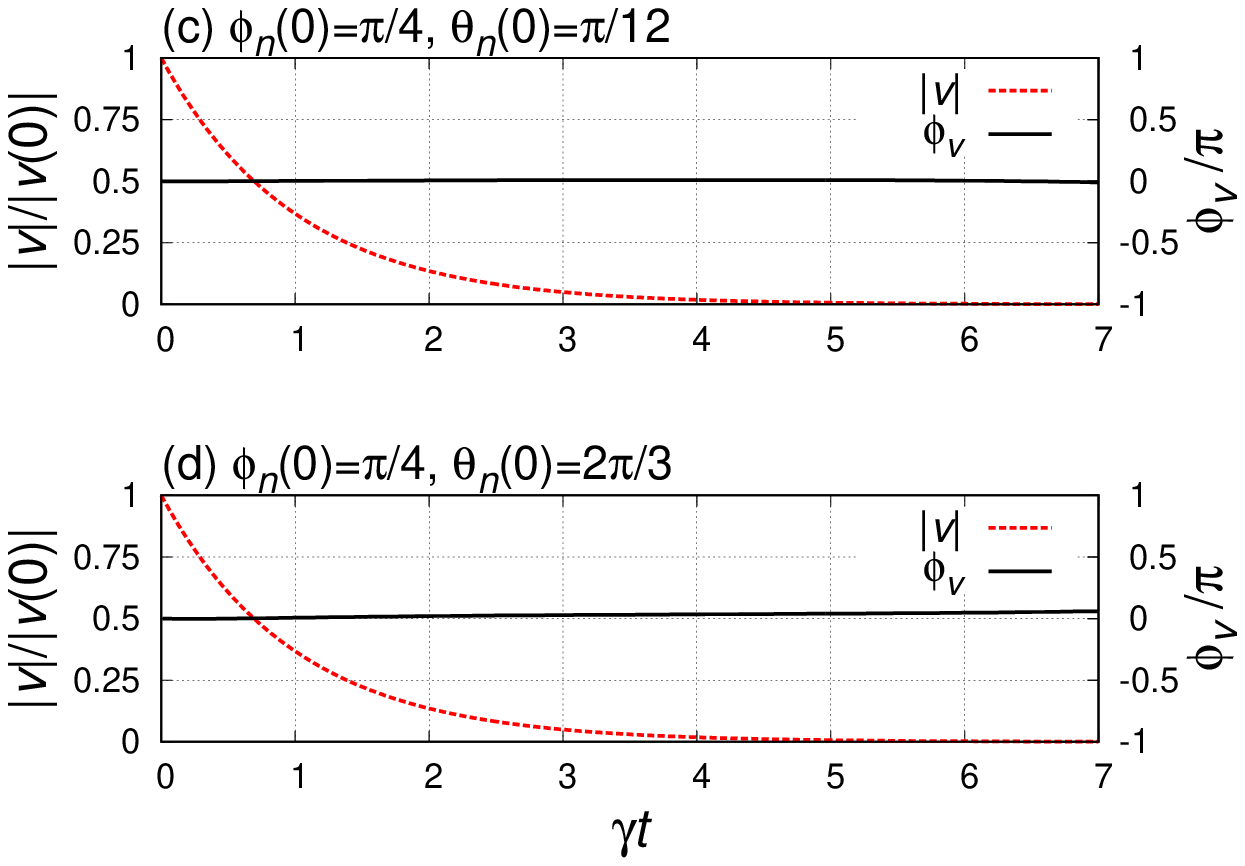}
    \end{tabular}
        \caption{
        Time evolution of the velocity, $\bm{v}=\dot{\bm{x}}$, at $B=0$, as plotted for
       $\tau_{\rm prec} \ll \tau_{\rm damp}$ ((a), (b))
      and $\tau_{\rm prec} \gg \tau_{\rm damp}$ ((c), (d)) under the initial conditions $\ltk \phi_n(0),\theta_n(0)\rtk=\ltk \pi/4, \pi/12\rtk$.
     $\phi_v$ denotes the angle between $\bm{v}(0)$ and $\bm{v}(t)$ in $x$-$y$ plane.
      The values $p_0$ and $\alpha$ are the same as in Fig.~\ref{fig3}.}
    \label{fig11}
\end{figure*}

\begin{figure*}[htbp]
    \begin{tabular}{rr}
      \includegraphics[width=0.5\linewidth]{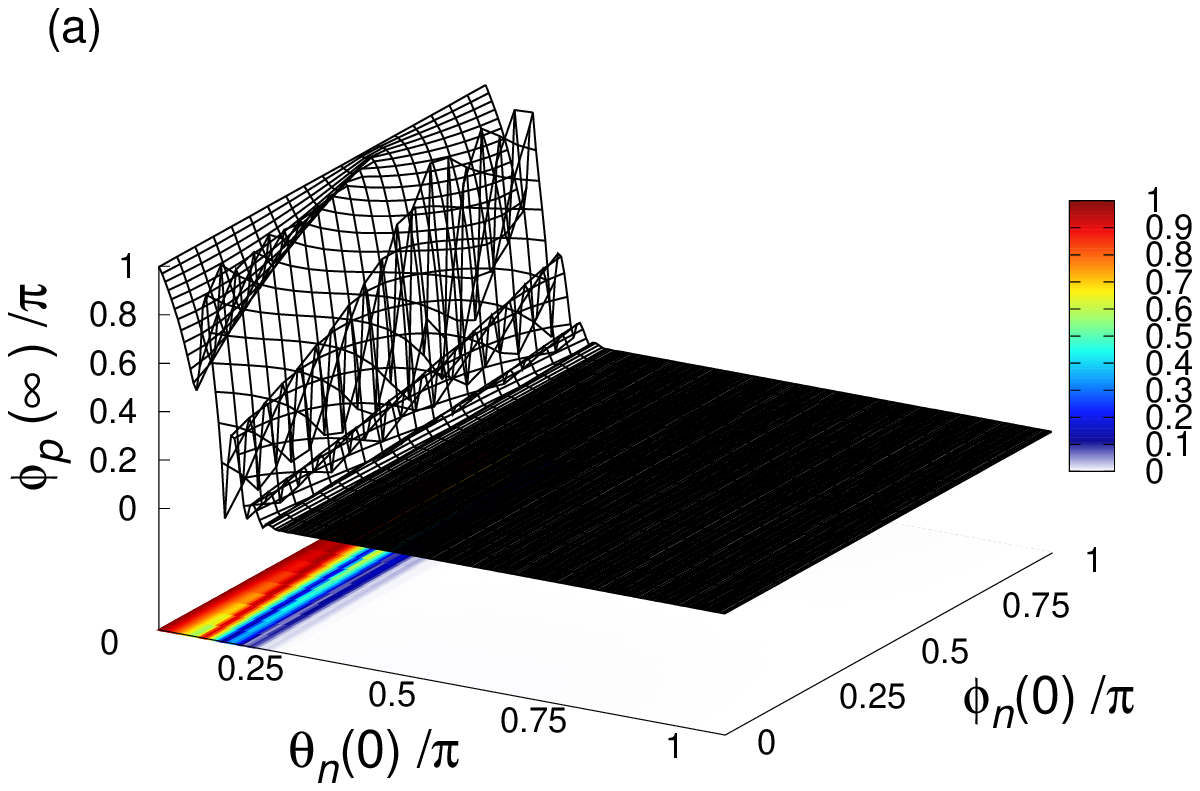} &
      \includegraphics[width=0.5\linewidth]{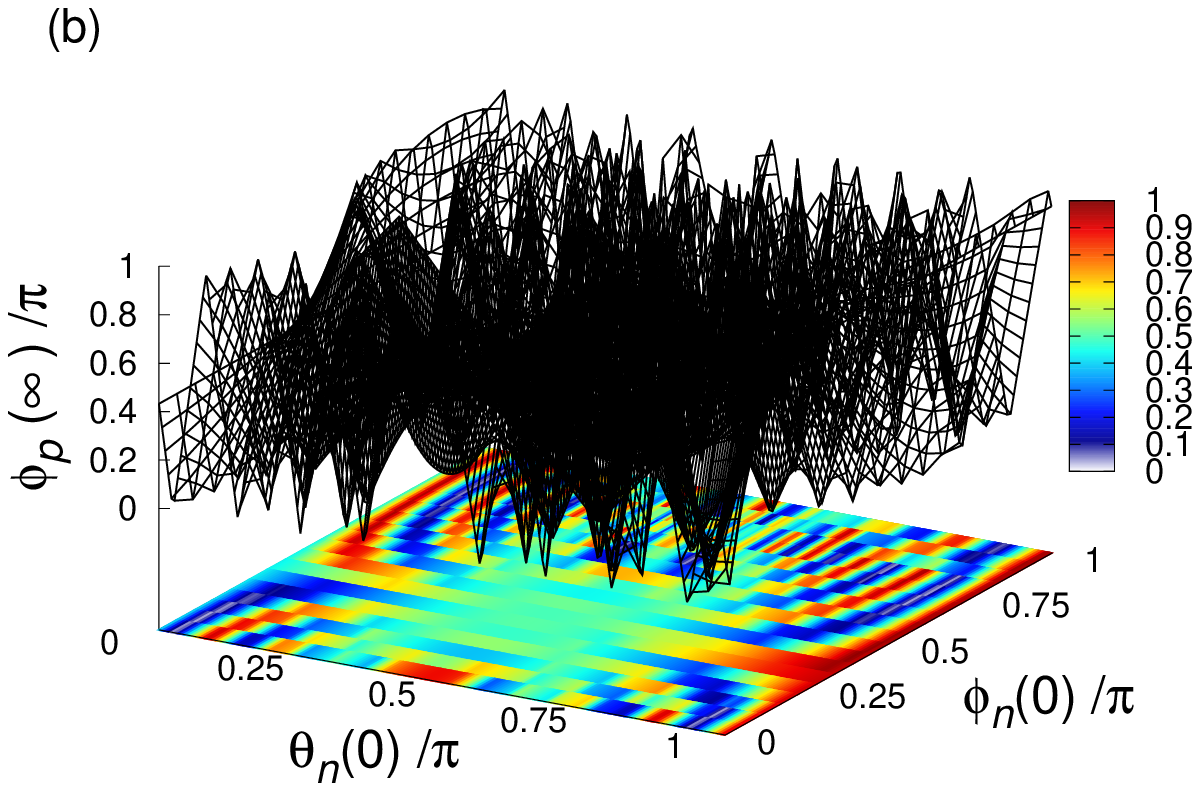}
    \end{tabular}
      \caption{Asymptotic values of the momentum angle $\phi_p(\infty)$ as a function of the initial values of the spin angles
      $\theta_n(0)$ and $\phi_n(0)$, as plotted for $\tau_{\rm prec} \ll \tau_{\rm damp}$ (a)
      and $\tau_{\rm prec} \gg \tau_{\rm damp}$ (b) at $B=0.027563\, \gamma<B_{\rm c}$.}
    \label{fig8}
\end{figure*}

\begin{figure*}[htbp]
  \centering
    \begin{tabular}{rr}
      \includegraphics[width=0.5\linewidth,height=5.5cm]{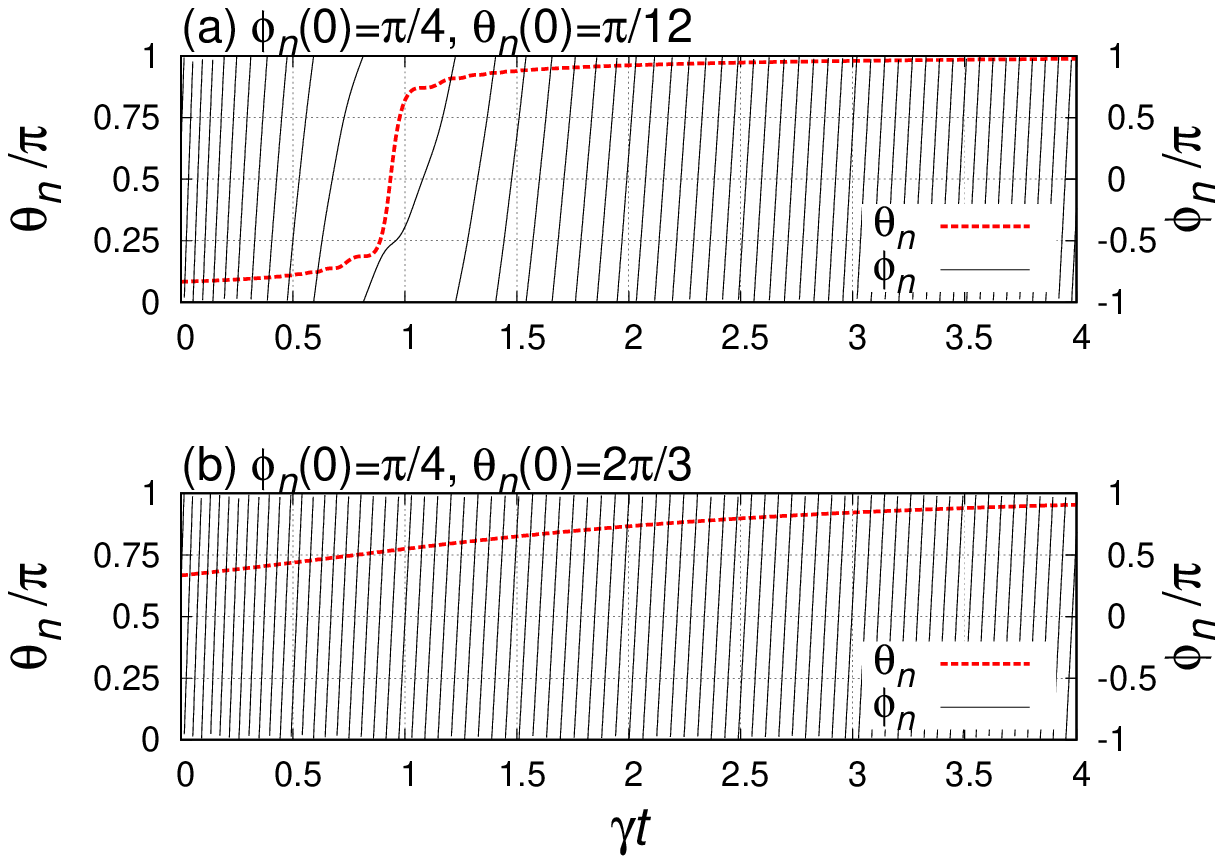} &
      \includegraphics[width=0.5\linewidth,height=5.5cm]{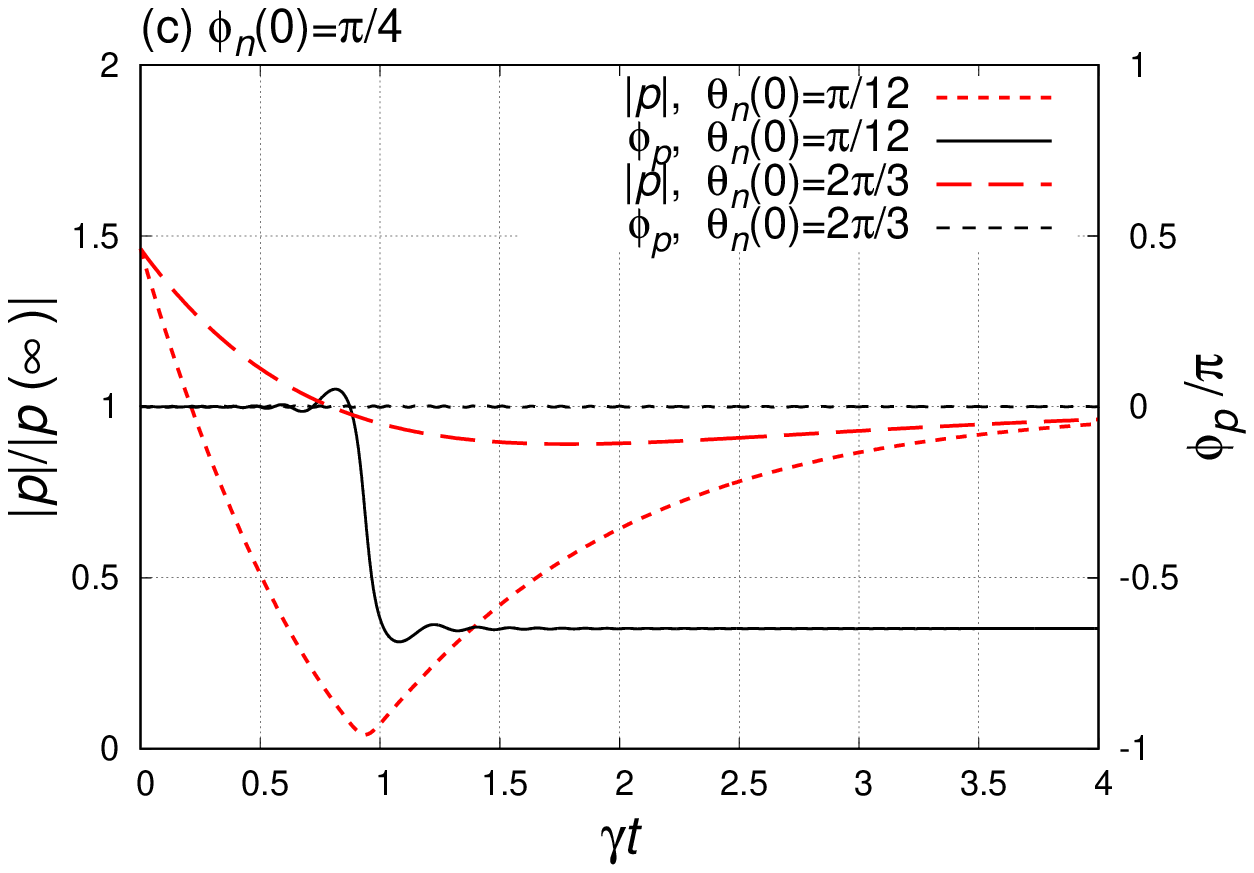} \\
      \vspace{0.1cm}\quad \\
      \includegraphics[width=0.5\linewidth,height=5.5cm]{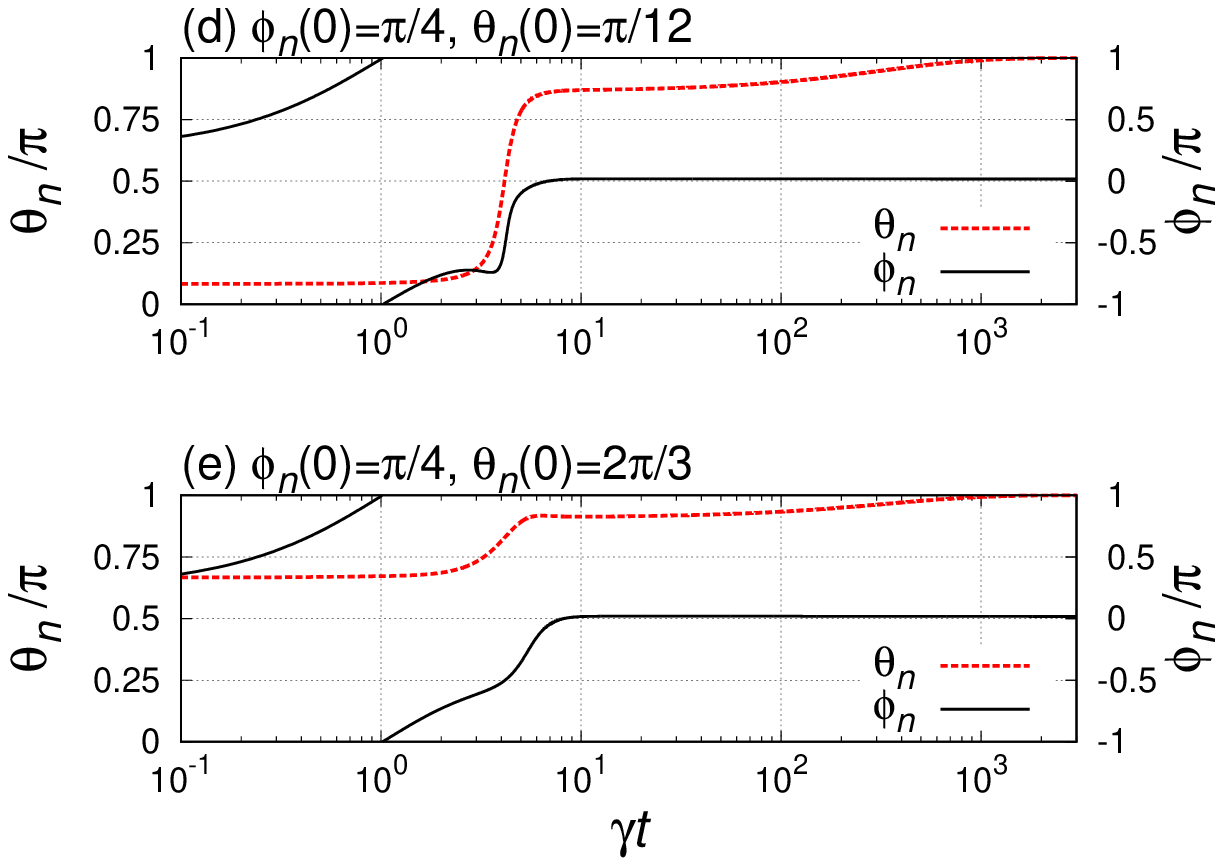} &
      \includegraphics[width=0.5\linewidth,height=5.5cm]{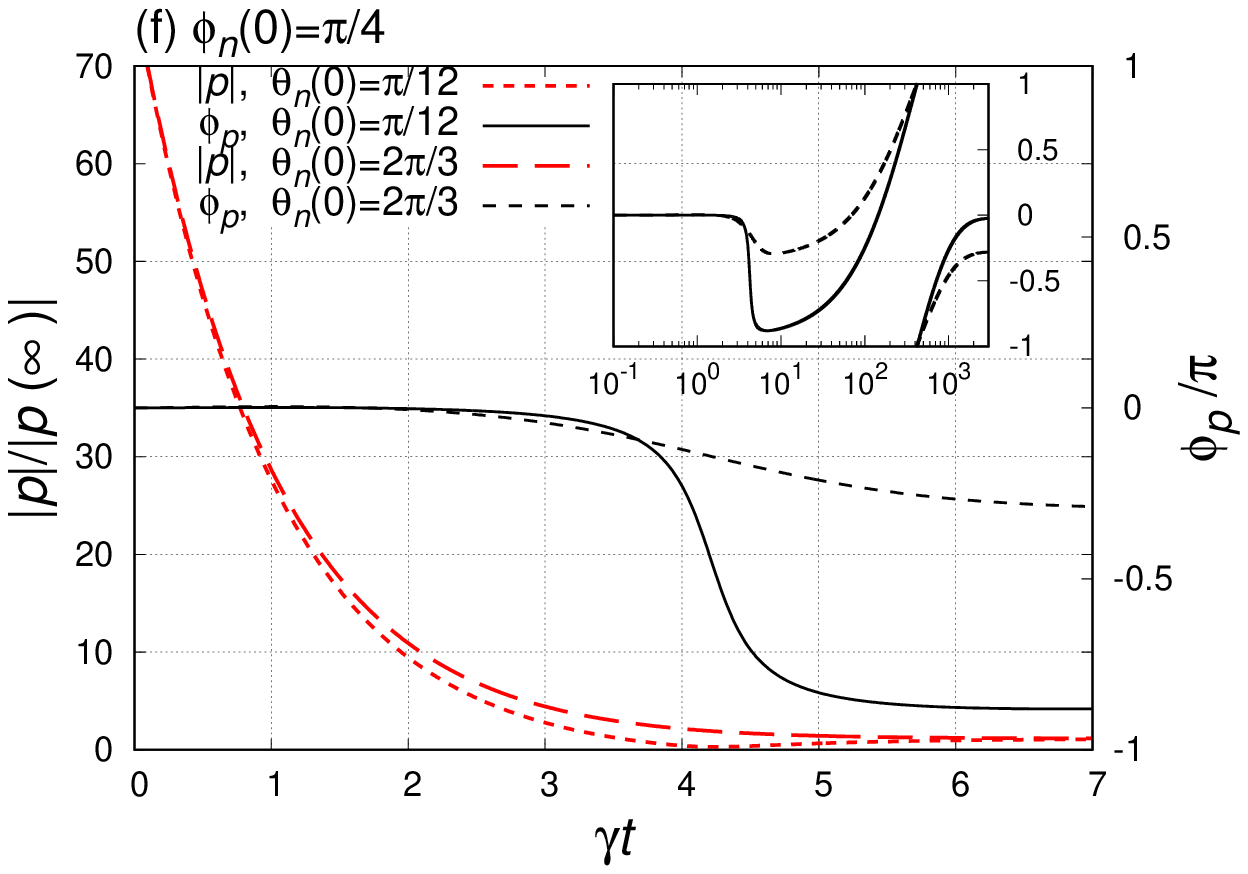}
    \end{tabular}
      \caption{
      Time evolution of the spin and monemtum  at $B=0.027563\, \gamma<B_{\rm c}$, calculated
     under the initial conditions $\ltk \phi_n(0), \theta_n(0) \rtk = \ltk \pi/4, \ltk \pi/12,2\pi/3 \rtk \rtk$.
   Panels (a), (b), (c) depict the results for $\tau_{\rm prec} \ll \tau_{\rm damp}$, and
     (d), (e), (f) for $\tau_{\rm prec} \gg \tau_{\rm damp}$. }
    \label{fig9}
\end{figure*}

\begin{figure*}[htbp]
  \centering
    \begin{tabular}{rr}
      \includegraphics[width=0.5\linewidth,height=5.5cm]{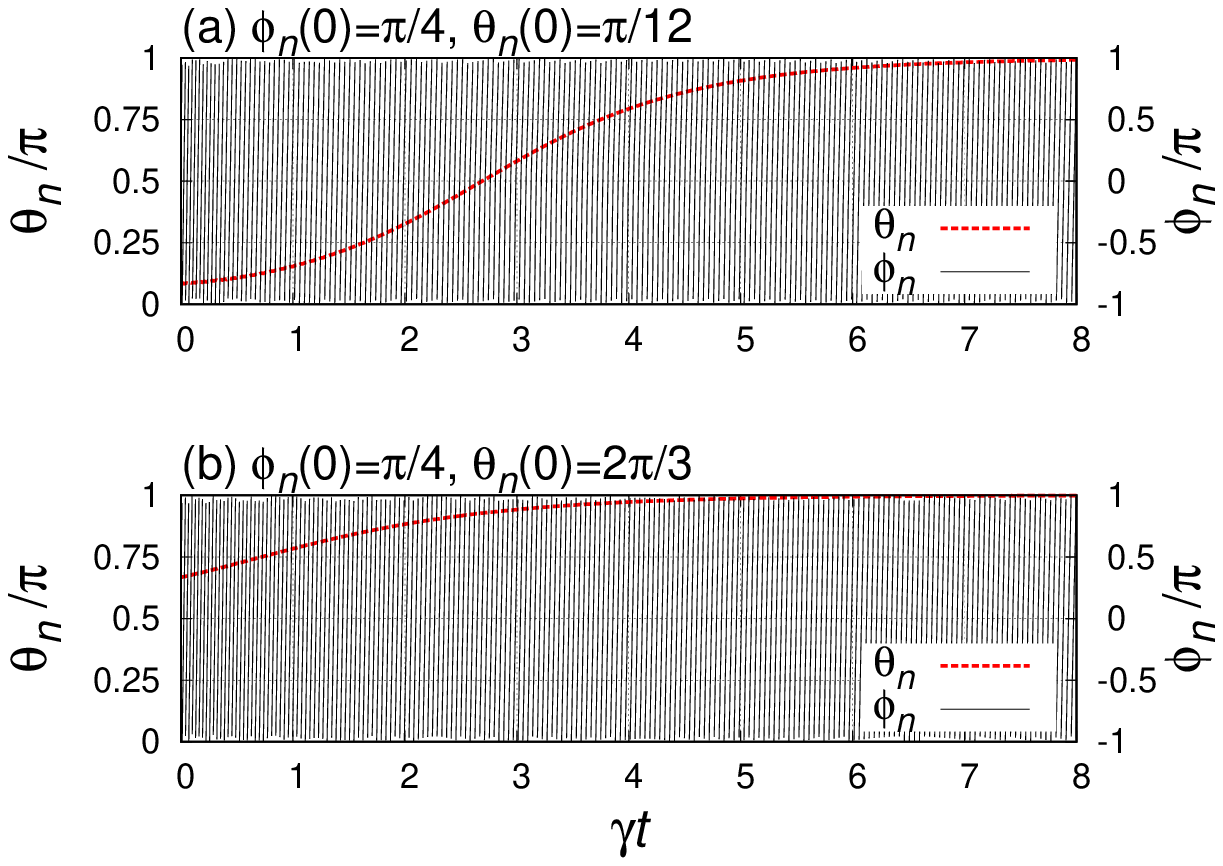} &
      \includegraphics[width=0.5\linewidth,height=5.5cm]{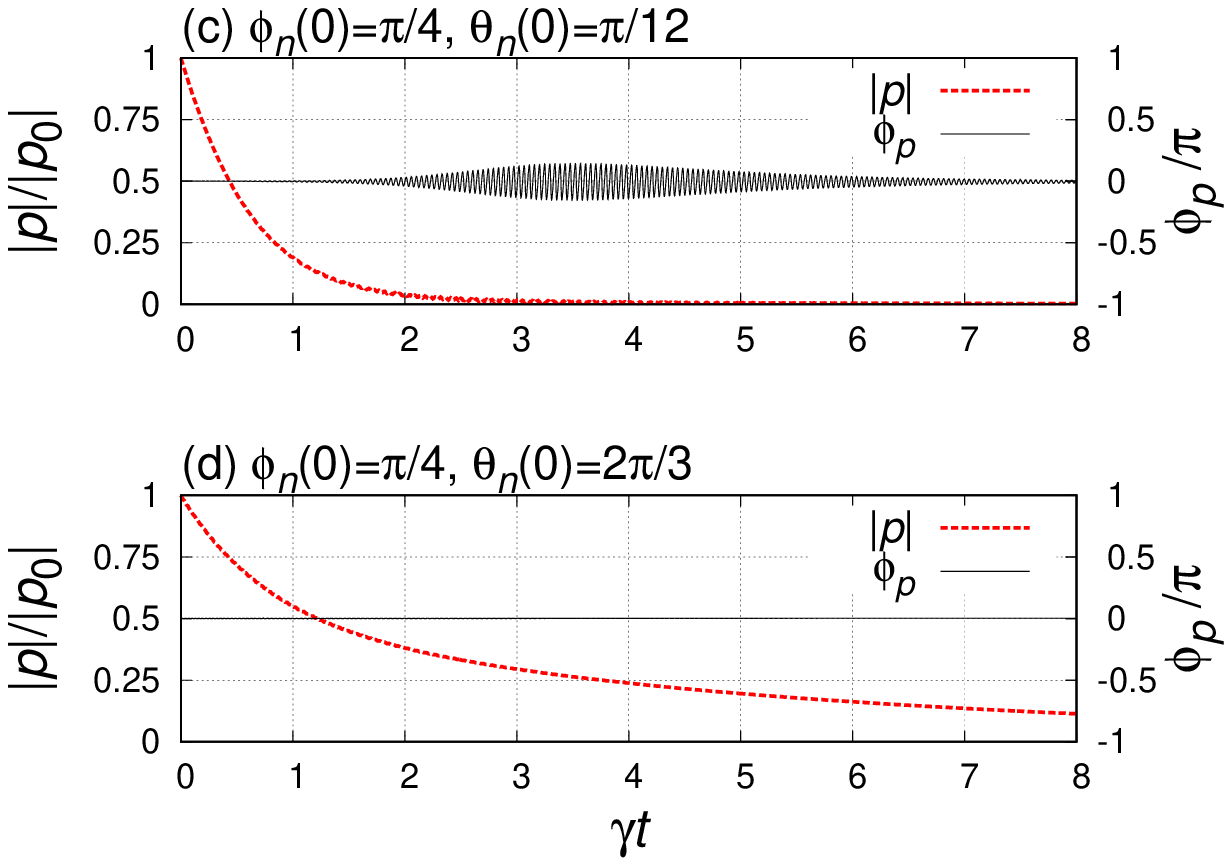} \\
      \vspace{0.1cm}\quad \\
      \includegraphics[width=0.5\linewidth,height=5.5cm]{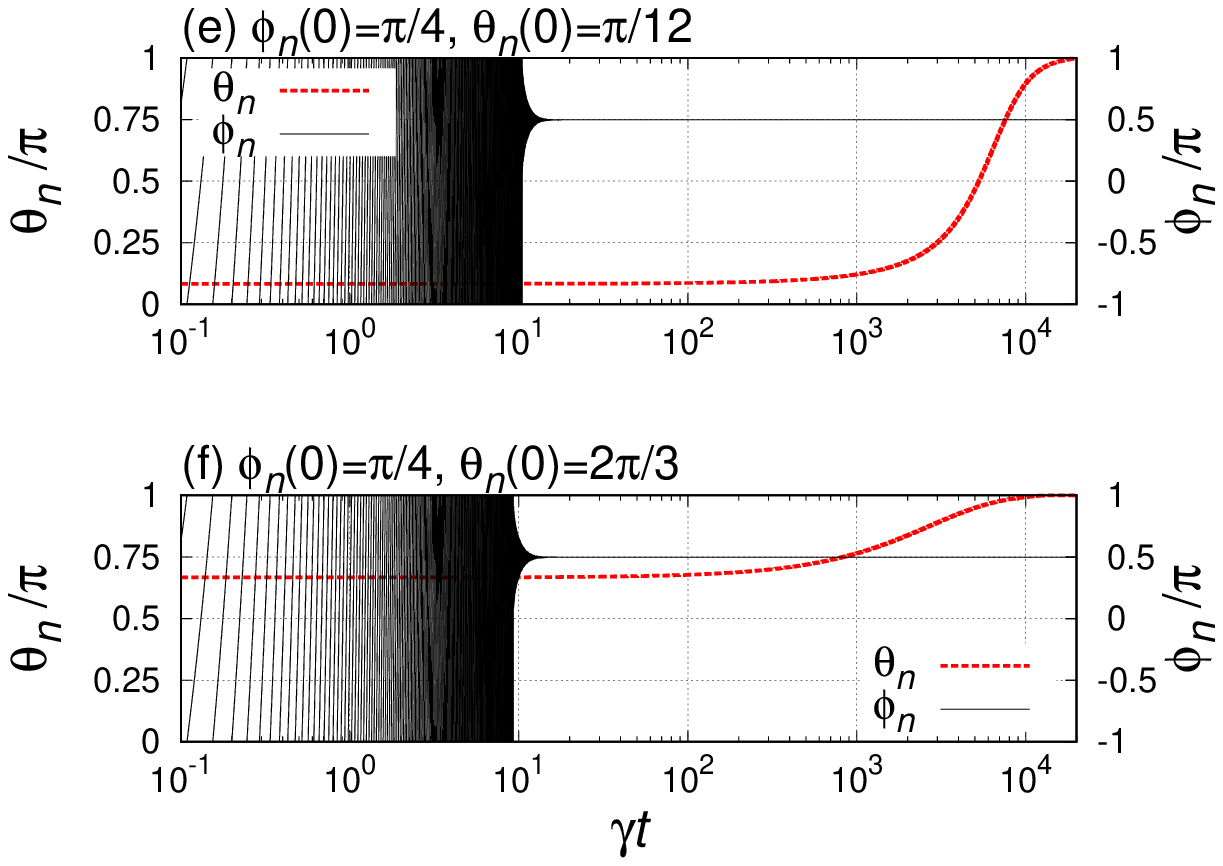} &
      \includegraphics[width=0.5\linewidth,height=5.5cm]{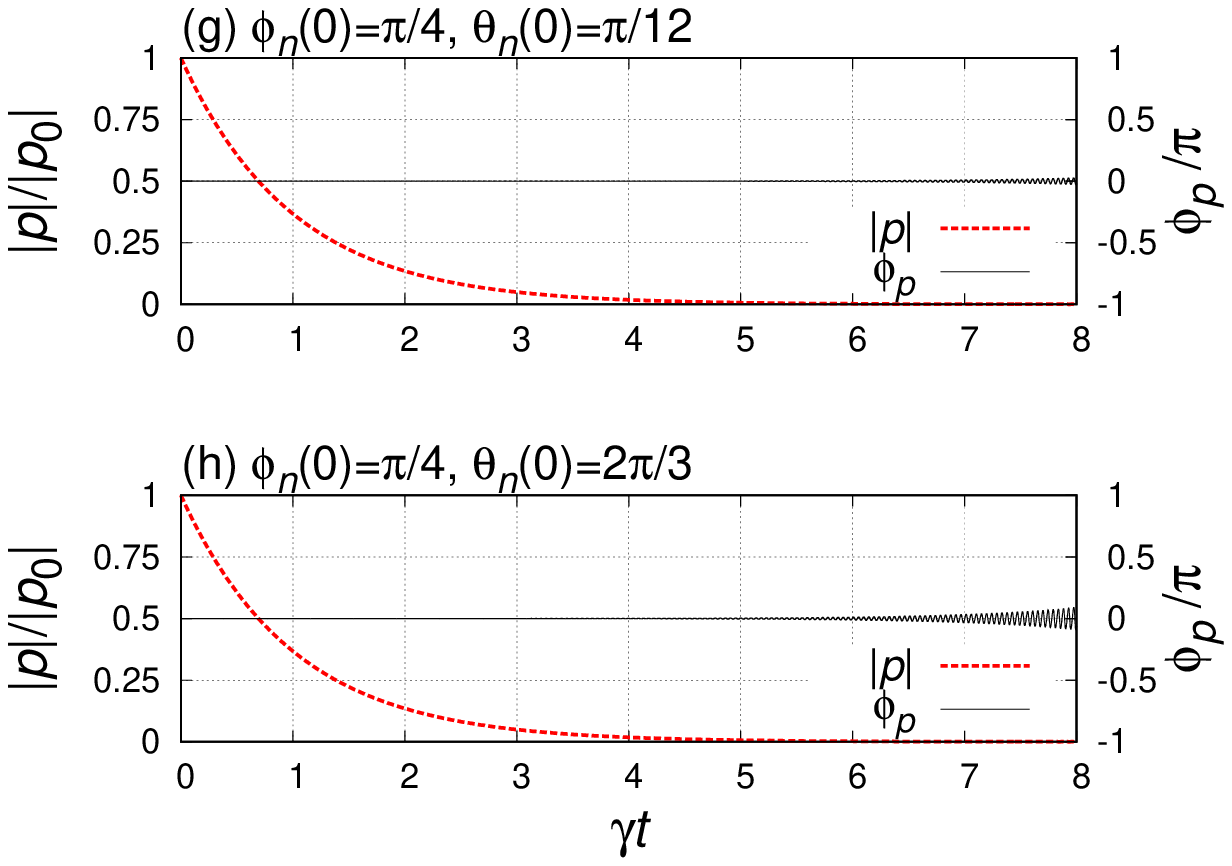}
    \end{tabular}
      \caption{
     Time evolution of the spin and monemtum at $B= 137.42 \gamma > B_{\rm c}$, calculated
     under the initial conditions $\ltk \phi_n(0), \theta_n(0) \rtk = \ltk \pi/4, \ltk \pi/12,2\pi/3 \rtk \rtk$.
    Panels (a), (b), (c), (d) depict the results for $\tau_{\rm prec} \ll \tau_{\rm damp}$, and
     (e), (f), (g), (h) for $\tau_{\rm prec} \gg \tau_{\rm damp}$.
     }
    \label{fig10}
\end{figure*}

In Sec.~III, we have presented numerical results only in the case of $\phi_n(0) = \pi/4$ and $B=0$,
which are in this appendix supplemented as follows:
Figs.~\ref{fig6}--\ref{fig7} show results for $\phi_n(0) = 0, \pi/2$ at $B=0$,
and Figs.~\ref{fig8}--\ref{fig10} for $\phi_n(0) = 0,\pi/4,\pi/2$ at $B\neq 0$.
The values of $\alpha$ and $\bp_0$ are the same as in Fig.~\ref{fig3}.
We also present the time evolution of the velocity in Fig.~\ref{fig11},
where the initial condition and parameter values are the same as in Fig.~\ref{fig4}.
\onecolumngrid
\end{document}